\documentclass[bibyear]{aa}
\usepackage{rotating} 
\usepackage{graphicx}
\usepackage{txfonts}
\begin{document}

\title{Spectroscopic variability of two Oe stars}
\author{G.\ Rauw\inst{1,2} \and T.\ Morel\inst{1,2} Y.\ Naz\'e\inst{1}\fnmsep\thanks{Research Associate FRS-FNRS (Belgium)} \and T.\ Eversberg\inst{2,3} \and F.\ Alves\inst{2} \and W.\ Arnold\inst{2} \and T.\ Bergmann\inst{2} \and N.G.\ Correia Viegas\inst{2} \and R.\ Fahed\inst{2,4} \and A.\ Fernando\inst{2} \and J.N. Gonz\'alez-P\'erez\inst{5} \and L.F.\ Gouveia Carreira\inst{2} \and A.\ Hempelmann\inst{5} \and T.\ Hunger\inst{2} \and J.H.\ Knapen\inst{2,6,7} \and R.\ Leadbeater\inst{2} \and F.\ Marques Dias\inst{2} \and M.\ Mittag\inst{5} \and A.F.J.\ Moffat\inst{2,4} \and N.\ Reinecke\inst{2} \and J.\ Ribeiro\inst{2} \and N.\ Romeo\inst{2} \and J. S\'anchez Gallego\inst{2,6,8} \and E.M.\ Dos Santos\inst{2} \and L.\ Schanne\inst{2} \and J.H.M.M.\ Schmitt\inst{5} \and K.-P.\ Schr\"oder\inst{9} \and O.\ Stahl\inst{2,10} \and Ba.\ Stober\inst{2} \and Be.\ Stober\inst{2} \and K.\ Vollmann\inst{2,3}}
\offprints{G.\ Rauw}
\mail{rauw@astro.ulg.ac.be}
\institute{Groupe d'Astrophysique des Hautes Energies, Institut d'Astrophysique et de G\'eophysique, Universit\'e de Li\`ege, All\'ee du 6 Ao\^ut, B\^at B5c, 4000 Li\`ege, Belgium
\and Mons Pro-Am collaboration
\and Schn\"orringen Telescope Science Institute, Waldbr\"ol, Germany
\and D\'epartement de Physique, Universit\'e de Montr\'eal, and Centre de Recherche en Astrophysique du Qu\'ebec, Canada
\and Hamburger Sternwarte, Universit\"at Hamburg, Germany
\and Instituto de Astrof\'{\i}sica de Canarias, E-38200 La Laguna, Tenerife, Spain
\and Departamento de Astrof\'{\i}sica de Canarias, Universidad de La Laguna, E-38205 La Laguna, Tenerife, Spain
\and Department of Physics \& Astronomy, University of Kentucky, Lexington, USA
\and Departamento de Astronom\'{\i}a, Universidad de Guanajuato, Guanajuato, Mexico
\and
Zentrum f\"ur Astronomie der Universit\"at Heidelberg, Landessternwarte K\"onigstuhl, Heidelberg, Germany} 
\date{Received date / Accepted date}
\abstract{The two Oe stars HD~45\,314 and HD~60\,848 have  recently been found to exhibit very different X-ray properties: whilst HD~60\,848 has an X-ray spectrum and the emission level typical of most OB stars, HD~45\,314 features a much harder and brighter X-ray emission, making it a so-called $\gamma$~Cas analogue.}{Monitoring the optical spectra could provide hints towards the origin of these very different behaviours.}{We analyse a large set of spectroscopic observations of HD~45\,314 and HD~60\,848, extending over 20 years. We further attempt to fit the H$\alpha$ line profiles of both stars with a simple model of emission line formation in a Keplerian disk.}{Strong variations in the strengths of the H$\alpha$, H$\beta$, and He\,{\sc i} $\lambda$\,5876 emission lines are observed for both stars. In the case of HD~60\,848, we find a time lag between the variations in the equivalent widths of these lines, which is currently not understood. The emission lines are double peaked with nearly identical strengths of the violet and red peaks. The H$\alpha$ profile of this star can be successfully reproduced by our model of a disk seen under an inclination of $30^{\circ}$. In the case of HD~45\,314, the emission lines are highly asymmetric and display strong line profile variations. We find a major change in behaviour between the 2002 outburst and the one observed in 2013. This concerns both the relationship between the equivalent widths of the various lines and their morphologies at maximum strength (double-peaked in 2002 versus single-peaked in 2013). Our simple disk model fails to reproduce the observed H$\alpha$ line profiles of HD~45\,314.}{Our results further support the interpretation that Oe stars do have decretion disks similar to those of Be stars. Whilst the emission lines of HD~60\,848 are explained well by a disk with a Keplerian velocity field, the disk of HD~45\,314 seems to have a significantly more complex velocity field that could be another signature of the phenomenon that produces its peculiar X-ray emission.}
\authorrunning{G. Rauw et al.}
\titlerunning{Spectroscopic variability of two Oe stars}
\maketitle
\section{Introduction \label{intro}}
By definition, Be stars are non-supergiant B stars that have at least once displayed Balmer line emission in their spectrum. Be stars are usually considered to be very rapidly rotating and non-radially pulsating B stars surrounded by an outwardly diffusing Keplerian disk\footnote{In addition to the `e' qualifier, several variants such as `(e)', `pe', `[e]' or `e+' have been introduced. See Table\,3 of Sota et al.\,(\cite{Sota}) for a detailed definition of these variants.}. About one fifth of the Galactic B stars belong to this category, but the strongest incidence of the Be phenomenon is found in the B1-2 subtypes (for recent reviews on Be stars, see Neiner \cite{Neiner}, Rivinius et al.\ \cite{Rivinius}).

Conti \& Leep (\cite{CL}) identified a small group of O-type stars - the so-called Oe stars - that seem to form an extension of the Be phenomenon towards higher stellar temperatures. Indeed, these stars display emission lines of the H\,{\sc i} Balmer series, as well as of other elements, such as He\,{\sc i} and Fe\,{\sc ii}, but do not exhibit conventional Of emission lines, such as He\,{\sc ii}\,$\lambda$\,4686 and N\,{\sc iii}\,$\lambda$\,4634-40. Negueruela et al.\ (\cite{NSB}) note that previous spectral classifications of Oe stars were probably too early because of infilling of the photospheric He\,{\sc i} absorptions by circumstellar emission. These authors thus argued that most Oe stars have true spectral types in the range O9 -- B0, with the notable exceptions of the O7.5\,IIIe star \object{HD~155\,806} (Negueruela et al.\ \cite{NSB}) and \object{HD~39\,680,} which was assigned an O6\,V:[n]pe var spectral type by Sota et al.\ (\cite{Sota}). 

Whilst the disks of some Be stars have been resolved in optical long baseline interferometry (see Rivinius et al.\ \cite{Rivinius}, and references therein), the more distant and hence fainter Oe stars are beyond the capabilities of current interferometry facilities. Alternatively, the presence of a disk-like outflow can also be inferred from the depolarization of the continuum polarization due to line emission. The lack of such a signature in some Oe stars led Vink et al.\ (\cite{Vink}) to question the existence of disk-like structures around these objects. Indeed, their linear spectropolarimetric observations revealed a clear depolarization effect across the H$\alpha$ emission line only in one case out of four Oe stars with H$\alpha$ emission\footnote{It must be stressed, though, that some other objects with known large-scale structures in their winds (e.g.\ $\theta^1$\,Ori C) were not detected either (see also Naz\'e \& Rauw \cite{NR}).}. 

The spectra of Be stars are strongly variable over a broad range of timescales. For instance, the relative strength of the peaks of double-peaked emission lines varies in so-called V/R (violet-to-red) irregular cycles that can last between a few weeks to decades with a mean around seven years (Neiner \cite{Neiner}, Rivinius et al.\ \cite{Rivinius}). The V/R cycles are usually attributed to precessing one-armed density waves in the disk (Okazaki \cite{Okazaki}). The precession periods of these density waves are two orders of magnitude longer than the orbital periods of the particles in the disk. Even more spectacular, the disk as a whole may dissipate and form anew (e.g.\ Draper et al.\ \cite{Draper}). 

With the exception of our previous study of the targets of the present paper and some scarce reports based on a limited number of data (Rauw et al.\ \cite{ibvs}, Copeland \& Heard \cite{Copeland}, Divan et al.\ \cite{Divan}, Gamen et al.\ \cite{Gamen}), the variability of Oe stars has received little attention so far. In the present paper, we study the variability of the emission lines in the spectra of \object{HD~45\,314} (PZ\,Gem) and \object{HD~60\,848} (BN\,Gem) classified as B0\,IVe and O9.5\,IVe, respectively, by Negueruela et al.\ (\cite{NSB})\footnote{In the work of Conti \& Leep (\cite{CL}), HD~45\,314 and HD~60\,848 were classified as O9?pe and O8\,V?pe, respectively. More recently, Sota et al.\ (\cite{Sota}) have assigned spectral types O9:\,npe and O8:\,V:\,pe to HD~45\,314 and HD~60\,848, respectively. Since Conti \& Leep's `?' and Sota et al.'s `:' qualifiers both mean uncertainties in the spectral subtype, both papers actually give the same spectral type for HD~60\,848 and only differ in the one for HD~45\,314 in the `n' qualifier (which means that the lines are broadened).} on various time scales from a few days up to several decades. Vink et al.\ (\cite{Vink}) report a spectropolarimetric detection of a disk in the case of HD~45\,314, but not for HD~60\,848. 

Both stars were recently studied in X-rays with {\it XMM-Newton}, revealing very contrasting X-ray properties (Rauw et al.\ \cite{Oeletter}). On the one hand, HD~60\,848 was found to display an X-ray emission similar to that of normal O-type stars, in terms of its X-ray luminosity and the temperature of the hot plasma. On the other hand, HD~45\,314 exhibits an unusually hard and bright X-ray emission that makes it a new member of the category of so-called $\gamma$~Cas analogues. The $\gamma$~Cas analogues are defined by their remarkable X-ray properties (Smith et al.\ \cite{Smith1}), which include a ratio $\frac{L_{\rm X}}{L_{\rm bol}} \sim 10^{-6}$, i.e.\ a factor 10 higher than for normal OB stars but at least an order of magnitude lower than for Be high-mass X-ray binaries, variations on timescales from minutes to months, and an X-ray spectrum that is dominated by a very hot plasma with $kT \geq 10$\,keV. The origin of the $\gamma$~Cas behaviour currently remains unknown, but the most promising scenarios include a wind-disk interaction (Smith et al.\ \cite{Smith1}) or perhaps accretion from the Be decretion disk onto a compact companion (White et al.\ \cite{White}).

The goal of the present study is to characterize the spectroscopic variability of these two Oe stars on a variety of timescales. For this purpose, we have conducted several dedicated observing campaigns, but we also retrieved a large number of spectra from various archives. 

\begin{table*}[t]
\caption{Journal of the observations used in this work.\label{Journal}}
\begin{center}
\begin{tabular}{c c c l c c c c c}
\hline
Instrument & Epoch & Resolving power & Wavelength domain & \multicolumn{2}{c}{Number of spectra} \\
           &       &                 &                   & HD~45\,314 & HD60848  \\
\hline
Aur\'elie @ 1.5m OHP & Feb.\ 1997 & 20000 & 6510 -- 6710\,\AA & 11 &  6 \\
                     & Nov.\ 1998 & 30000 & 6500 -- 6620\,\AA &  7 &  6 \\
                     & Nov.\ 1998 & 30000 & 4795 -- 4925\,\AA &  1 &  1 \\
                     & Sep.\ 2000 & 10000 & 4460 -- 4900\,\AA &  3 &  2 \\
                     & Sep.\ 2001 & 10000 & 6350 -- 6770\,\AA &  3 &  - \\
                     & Sep.\ 2001 & 10000 & 4460 -- 4900\,\AA &  1 &  - \\
                     & Jan.\ 2008 & 10000 & 6350 -- 6770\,\AA &  1 &  1 \\
                     & Jan.\ 2008 & 10000 & 5490 -- 5920\,\AA &  3 &  3 \\
                     & Sep.\ 2008 & 10000 & 4460 -- 4900\,\AA &  1 &  - \\
                     & Dec.\ 2009 & 20000 & 4450 -- 4670\,\AA &  1 &  5 \\
                     & Dec.\ 2010 & 20000 & 4450 -- 4670\,\AA &  - &  1 \\
FEROS @ 1.5m ESO     & May 1999   & 48000 & 3900 -- 7100\,\AA & 10 & 10 \\
                     & May 2000   & 48000 & 3900 -- 7100\,\AA &  4 &  6 \\
                     & May 2001   & 48000 & 3900 -- 7100\,\AA &  3 &  3 \\
                     & Mar.\ 2002 & 48000 & 3900 -- 7100\,\AA &  3 &  3 \\
FEROS @ 2.2m ESO     & Jan.-Feb.\ 2006 & 48000 & 3900 -- 7100\,\AA & 3 & - \\
RGO @ AAT            & Dec.\ 1994 &  3500 & 3700 -- 6000\,\AA &  1 &  1 \\
IDS @ INT            & Jan.\ 1998 &  4500 & 4620 -- 5020\,\AA &  - &  2 \\
                     & Oct.\ 2001 &  3500 & 3950 -- 4880\,\AA &  1 &  1 \\
                     & Jan.\ 2002 &  4500 & 6100 -- 7100\,\AA &  1 &  1 \\
UVES @ UT2           & Jan.\ 2001 & 54000 & 3280 -- 4560\,\AA &  - &  1 \\
                     & Jan.\ 2001 & 87000 & 4720 -- 6830\,\AA &  - &  1 \\
Elodie @ 1.9m OHP    & Nov.\ 2004 & 42000 & 3905 -- 6810\,\AA &  1 &  1 \\
LHIRES III @ Mons    & Dec.\ 2008 - Mar.\ 2009 &  3500 & 6360 -- 6950\,\AA & 35 & 40 \\
Sophie @ 1.9m OHP    & Mar.\ 2012 & 40000 & 3870 -- 6940\,\AA &  1 &  - \\
Coralie @ Euler & Mar.\ - April 2012 & 55000 & 3850 -- 6890\,\AA &  2 &  1 \\
FIES @ NOT           & Jan.\ 2011 & 46000 & 3700 -- 7300\,\AA &  1 &  2 \\
                     & Dec.\ 2012 & 46000 & 3700 -- 7300\,\AA &  1 &  1 \\
                     & Jan.\ 2013 & 46000 & 3700 -- 7300\,\AA &  3 &  - \\ 
HEROS @ TIGRE & Sep.\ 2013 - Feb.\ 2014 & 20000 & 3500 -- 8800\,\AA & 10 & 9 \\
\hline
\end{tabular}
\end{center}
\end{table*}

\section{Observations \label{obs}}
The data used in the present work come from a variety of observatories and equipments (see Table\,\ref{Journal} for details). The largest part of our data set was obtained via a collaboration between amateur and professional astronomers. Indeed, Be stars are prominent targets for amateur astronomer spectroscopists, and several collaborations have previously taken place between professional and amateur astronomers (e.g.\ Neiner \cite{Neiner}). In our case, most of the amateur data were taken during a four-month campaign (Eversberg \cite{Eversberg}) conducted at the 50\,cm Mons telescope located at the Observatorio del Teide and operated by the Instituto de Astrof\'{\i}sica de Canarias (IAC) in Tenerife. A LHIRES III spectrograph was installed on the Mons telescope and operated between early December 2008 and the end of March 2009. 

The main goal of this campaign was to monitor the colliding wind binary WR\,140 during its periastron passage (see Fahed et al.\ \cite{Fahed}). Since WR\,140 was observable for most of the run only close to either sunset or sunrise, additional targets were observed, including HD~45\,314 and HD~60\,848 (Morel et al.\ \cite{Morel}). The wavelength coverage was between 6360 and 6950\,\AA\ with a reciprocal dispersion of $\sim 0.34$\,\AA\,pixel$^{-1}$. The data were reduced using the IRAF software. To correct for some drifts in the instrumental set-up, we re-aligned the wavelength-calibrated spectra by means of the diffuse interstellar band at 6613.910\,\AA\  or by means of strong telluric lines (e.g.\ 6892.3\,\AA). After rejecting some spectra that were affected by strong ghosts due to remanence of the CCD after Ne-lamp exposures, we were left with 35 usable spectra of HD~45\,314 and 40 spectra of HD~60\,848. The HD~60\,848 time series provides a rather uniform coverage over the full duration of the campaign, whilst there is a gap in the middle of the campaign for the HD~45\,314 time series. A few additional amateur spectra of HD~60\,848 were retrieved from the BeSS database (Neiner \cite{Neiner}), operated at LESIA, Observatoire de Meudon, France\footnote{http://basebe.obspm.fr}. 

The rest of the data used in this paper were taken with professional instruments. Dedicated campaigns were conducted between February 1997 and September 2010 with the Aur\'elie spectrograph (Gillet et al.\ \cite{Gillet}) at the 1.52\,m telescope of the Observatoire de Haute Provence (OHP, France). Various set-ups (both in terms of wavelength coverage and resolving power) were used. Between May 1999 and March 2002 echelle spectra were taken with the FEROS instrument (Kaufer et al.\ \cite{Kaufer}) mounted on the 1.5\,m telescope at La Silla (ESO, Chile). The data were reduced using the MIDAS software along with private software tools designed for the reduction of spectroscopic data taken with these instruments. Part of the Aur\'elie and FEROS data (up until 2002) were already used in the study of Rauw et al.\ (\cite{ibvs}). Another dedicated campaign was conducted between September 2013 and February 2014 on the TIGRE telescope (formerly known as the Hamburg Robotic Telescope, Hempelmann et al.\ \cite{Hempelmann}, Mittag et al.\ \cite{Mittag}, Schmitt et al.\ \cite{Schmitt}) installed at La Luz Observatory (Guanajuato, Mexico). The TIGRE features the refurbished HEROS echelle spectrograph (Kaufer \cite{Kaufer2}) and is operated in a fully robotic way. We also include the echelle spectra taken with the Coralie and Sophie spectrographs, respectively, at the Swiss 1.2\,m Leonhard Euler Telescope (La Silla, Chile) and the 1.93\,m telescope at OHP, which we obtained in March and April 2012 to support our X-ray observations of the Oe stars (see Rauw et al.\ \cite{Oeletter}).

To complement our data, we also extracted spectra from a number of archives. We retrieved echelle spectra from the Fibre-fed Echelle Spectrograph (FIES) at the 2.56\,m Nordic Optical Telescope at the Observatorio del Roque de Los Muchachos (La Palma, Canary Islands, Spain). These data were reduced with the IRAF software. Long-slit spectra were retrieved from the archives of the Intermediate Dispersion Spectrograph (IDS) at the Cassegrain focus of the 2.5\,m Isaac Newton Telescope at the Observatorio del Roque de Los Muchachos and from the Royal Greenwich Observatory spectrograph at the 3.9\,m Anglo Australian Telescope at the Anglo Australian Observatory (Siding Spring, Australia). The long-slit spectra were reduced and calibrated using the MIDAS software. Finally, calibrated echelle spectra were taken from the archives of the Elodie instrument at the 1.93\,m OHP telescope (Moultaka et al.\ \cite{Moultaka}), as well as from the UVES spectrograph at the ESO Very Large Telescope (Cerro-Paranal, Chile). 

For those data sets where we had observations of fast-rotating standard stars at different airmasses, we used the latter to build a template file of the telluric absorption lines. In all other cases, we used the {\tt telluric} tool within IRAF along with the list of telluric lines of Hinkle et al.\ (\cite{Hinkle}) to remove the telluric lines in the He\,{\sc i} $\lambda$\,5876 and H$\alpha$ regions. Given the strengths of the H$\alpha$ emissions we are dealing with, the impact of the telluric lines on the equivalent widths is rather modest, well below the uncertainties due to the normalization. The impact of the correction is somewhat greater for the weaker He\,{\sc i} $\lambda$\,5876 lines.

\begin{figure*}[t!hb]
\begin{center}
\resizebox{16cm}{!}{\includegraphics{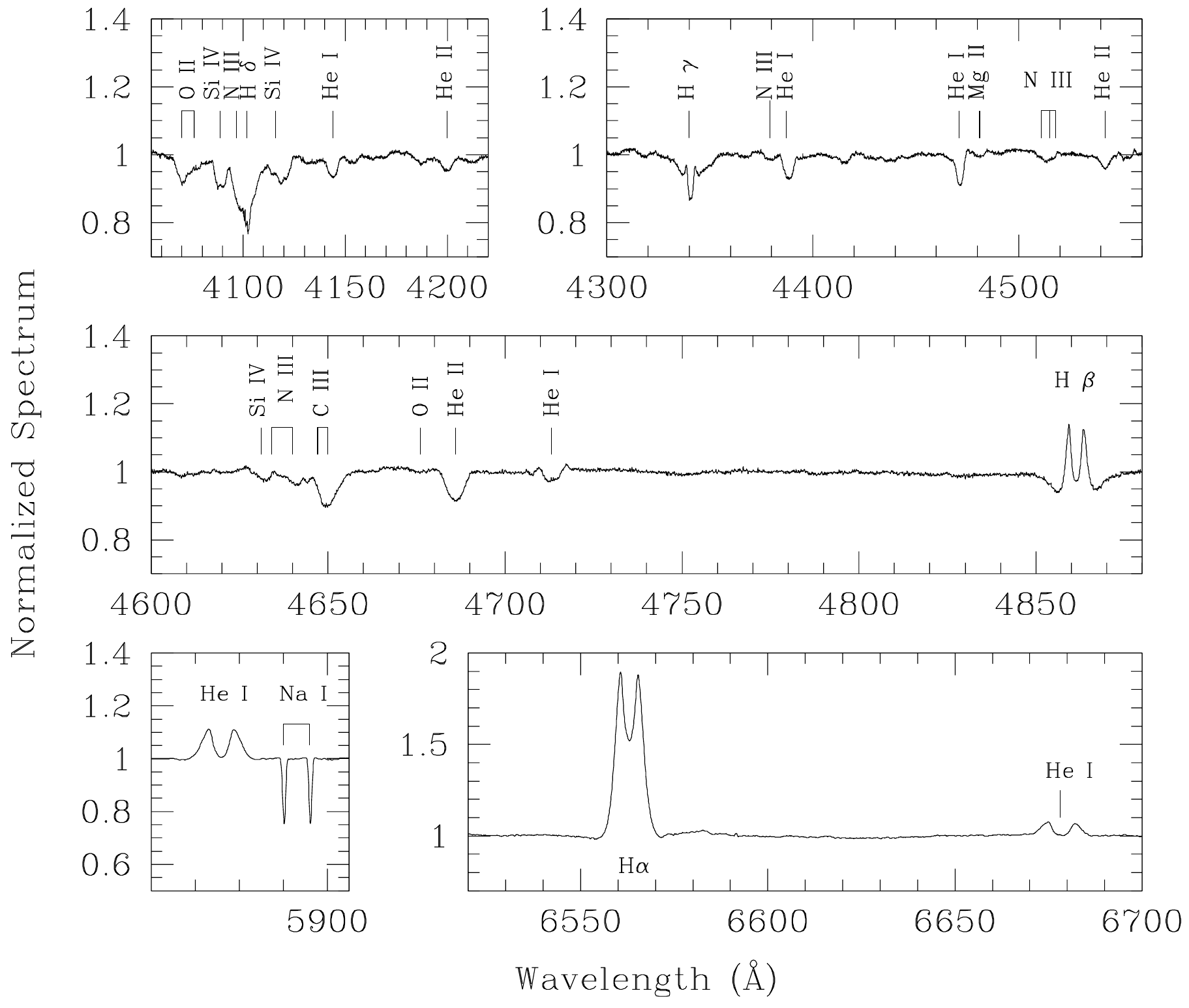}}
\end{center}
\caption{Important spectral regions of HD~60\,848 as observed with the HEROS spectrograph in December 2013 when the global emission level of the star was low.\label{spectrum60848}}
\end{figure*}
\begin{figure*}[t!hb]
\begin{center}
\resizebox{16cm}{!}{\includegraphics{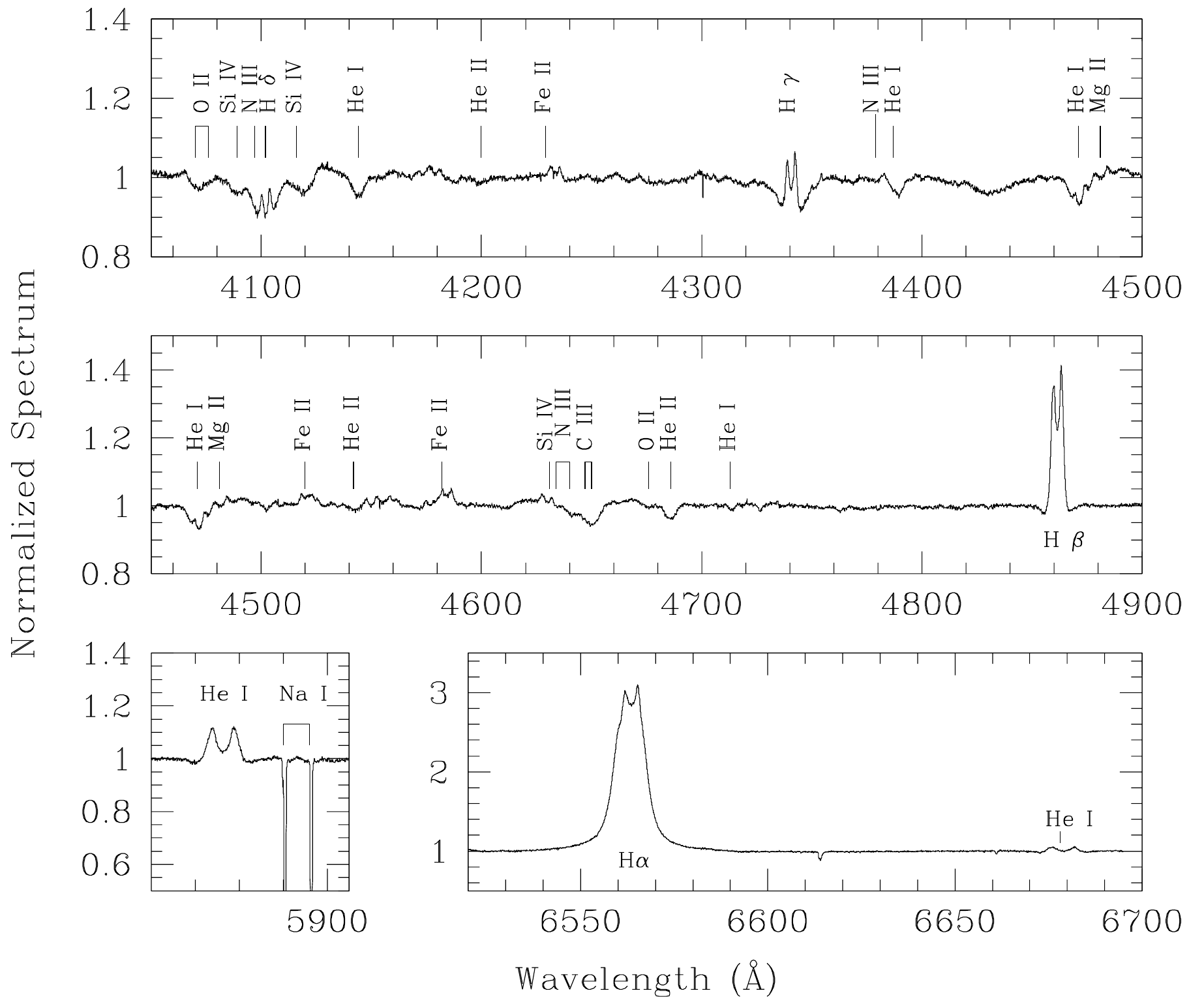}}
\end{center}
\caption{Same as Fig.\,\ref{spectrum60848} but for HD~45\,314 as observed with the FIES spectrograph in January 2011 when the global emission level of the star was low. \label{spectrum45314}}
\end{figure*}

Figure\,\ref{spectrum60848} illustrates some important wavelength domains in the spectrum of HD~60\,848 at a time when the circumstellar emission was at a low level. Beside the Balmer and He\,{\sc i} circumstellar emission lines, we note the presence of a number of photospheric absorption lines, associated with He\,{\sc ii}, Si\,{\sc iv,} and N\,{\sc iii}, which are typically observed in late-type O-stars. Figure\,\ref{spectrum45314} provides the same information for HD~45\,314. In addition to these lines, the spectrum of the latter star also displays a number of Fe\,{\sc ii} emission lines. The lack of a strong photospheric He\,{\sc ii} $\lambda$\,4542 absorption indeed indicates that the central star of HD~45\,314 must have a later spectral type than HD~60\,848. 

To evaluate the projected rotational velocity of our targets, we have extracted the line profile of the He\,{\sc ii} $\lambda$\,4686 absorption in the low-emission state (HEROS spectra from December 2013 for HD~60\,848 and FIES spectrum from January 2011 for HD~45\,314). Indeed, for both stars, He\,{\sc ii} $\lambda$\,4686 is the strongest absorption line that is free of emission from the circumstellar disk and is relatively free of blends with other lines. We then applied the Fourier transform method (Gray \cite{Gray}, Sim\'on-D\'{\i}az \& Herrero \cite{Simon-Diaz}) to derive $v\,\sin{i}$. For HD~60\,848 and HD~45\,314, we obtain $(230 \pm 5)$\,km\,s$^{-1}$ and $(210 \pm 10)$\,km\,s$^{-1}$, respectively. These values are in good agreement with those obtained by Penny (\cite{Penny}): 231\,km\,s$^{-1}$ for HD~60\,848 and 243\,km\,s$^{-1}$ for HD~45\,314, via a cross-correlation method of {\it IUE} high-resolution spectra. Our value of HD~45\,314 is slightly more uncertain because the blue wing of the He\,{\sc ii} $\lambda$\,4686 line is blended with O\,{\sc ii} $\lambda$\,4676. In both cases, the derived value of $v\,\sin{i}$ is likely a lower limit. Indeed, as a result of gravity darkening, the He\,{\sc ii} absorption most likely forms nearer to the hotter stellar poles and is less sensitive to the cooler equatorial regions that rotate faster.

\section{Measurements\label{measurements}}
Because we are dealing with a rather heterogeneous data set, combining data from different instruments and covering different wavelength domains at different spectral resolutions, we first need to decide what quantities we wish to measure. For the purpose of the present analysis, we focus on the radial velocities (RVs), equivalent widths (EWs), and violet over red (V/R) ratios\footnote{We define the V/R ratio as the ratio between the continuum-subtracted line intensities of the violet and red emission peaks.} of those spectral lines that are most often covered by our observations. For HD~60\,848, we obtain RVs for the He\,{\sc ii} $\lambda\lambda$\,4542, 4686, and 5412 absorption lines\footnote{We adopted the rest wavelengths 4541.59, 4685.68, and 5411.52\,\AA\ from Underhill (\cite{Underhill}).}. For HD~45\,314, we measure the RVs of the He\,{\sc ii} $\lambda$\,4686 absorption, as well as of the highest emission peak of the H$\alpha$ line. He\,{\sc ii} lines are intrinsically broader than\ He\,{\sc i} or metallic lines, for example, and are therefore subject to larger RV uncertainties. However, in the present case, the He\,{\sc ii} lines are in fact the only relatively strong and relatively isolated absorption lines that are not affected by emission from circumstellar material, and are thus suitable for RV measurements. For both stars, V/R ratios and EWs are measured for the H$\beta$, H$\alpha$, and He\,{\sc i} $\lambda\lambda$\,5876, 6678 lines. The EWs are measured over fixed wavelength windows. For HD~45\,314, we use 4850 -- 4875, 6530 -- 6600, 5865 -- 5885, and 6670 -- 6689\,\AA\ for the H$\beta$, H$\alpha$ and He\,{\sc i} $\lambda\lambda$\,5876, 6678 lines, respectively. For HD~60\,848, we adopt the same windows, except for the H$\alpha$ line where we instead use 6550 -- 6575\,\AA. The uncertainties on the EWs were estimated using the formalism of Vollmann \& Eversberg (\cite{VE}). 

\section{Results}
\subsection{HD~60\,848}
\begin{figure}[h!]
\begin{center}
\resizebox{9cm}{!}{\includegraphics{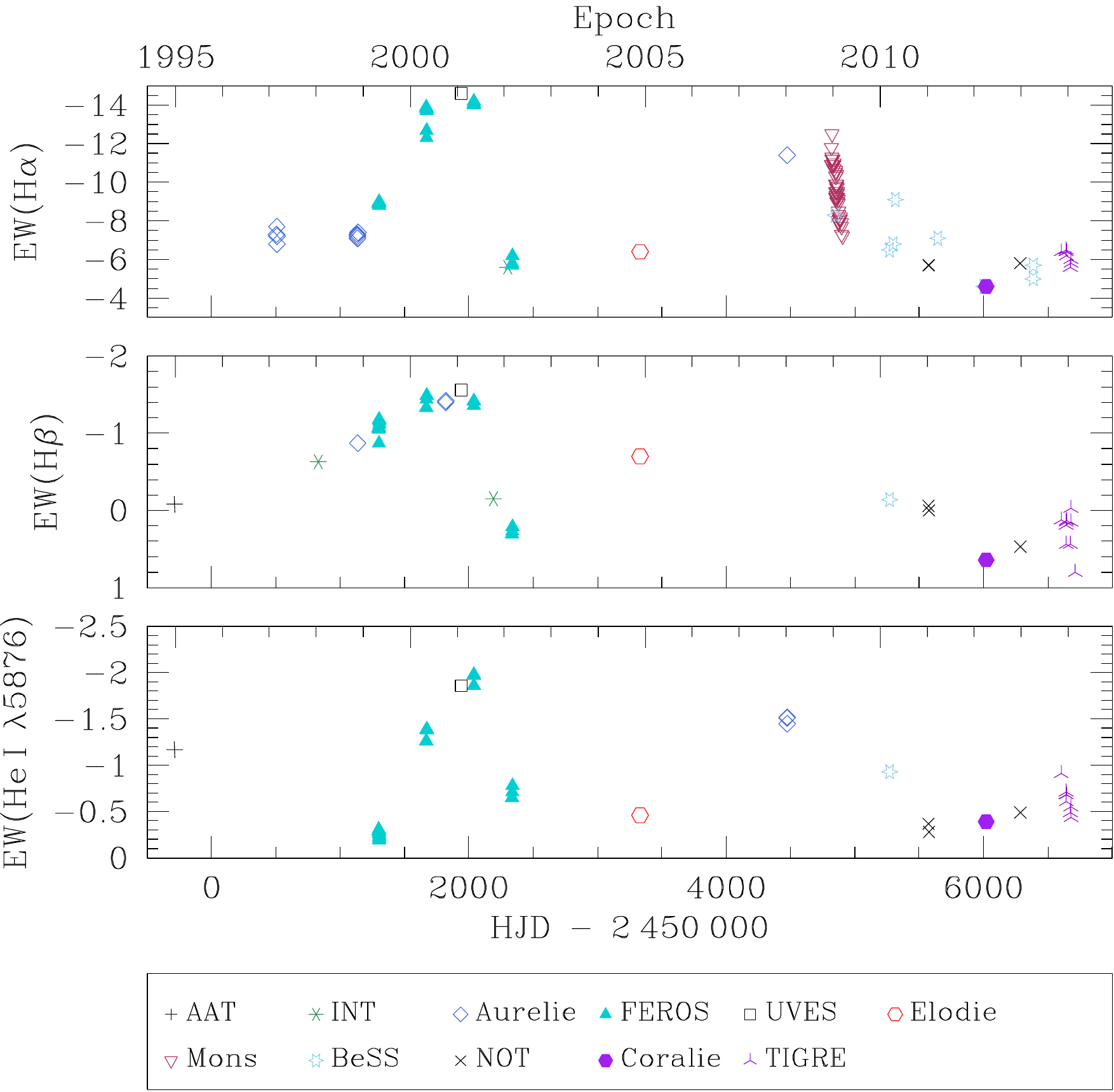}}
\end{center}
\caption{Long-term variations in the equivalent widths (in \AA) of the H$\alpha$, H$\beta$, and He\,{\sc i} $\lambda$\,5876 lines in the spectrum of HD~60\,848. The lower panel yields the meaning of the different symbols. For each of the three upper panels, the lower horizontal axis yields the date in HJD $ - 2\,450\,000$, whilst the upper horizontal axis yields the epoch expressed as Gregorian calendar years. Typical uncertainties on the EWs are 0.25, 0.25, and 0.20\,\AA\ for the H$\alpha$, H$\beta$, and He\,{\sc i} $\lambda$\,5876 lines, respectively. The Coralie spectrum was taken simultaneously with the {\it XMM-Newton} observation discussed by Rauw et al.\ (\cite{Oeletter}). \label{HD60848history}}
\end{figure}

\begin{figure}[h!]
\begin{center}
\resizebox{9cm}{!}{\includegraphics{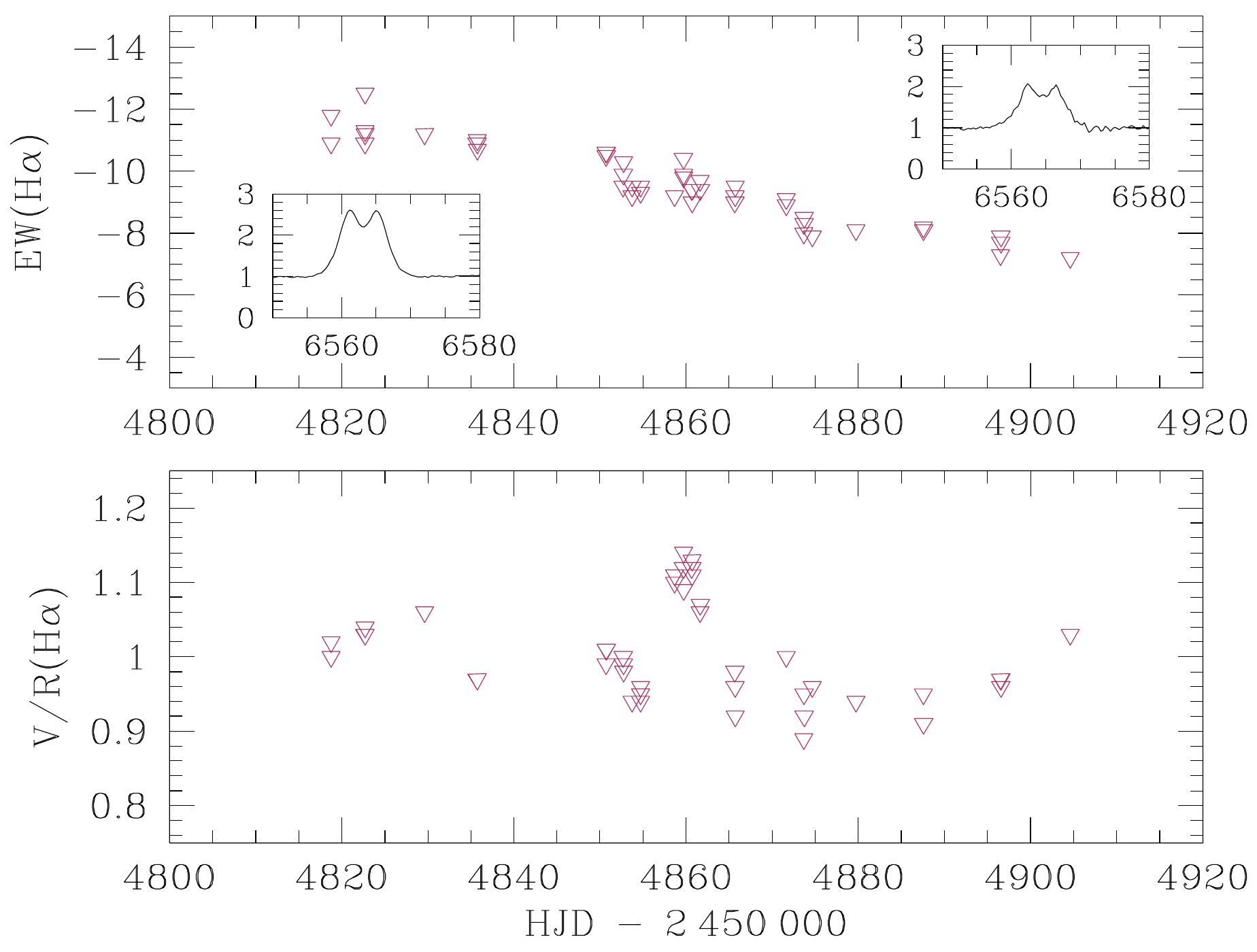}}
\end{center}
\caption{Zoom on the trends observed during the intense monitoring of the Mons campaign for the equivalent widths and V/R ratio of the H$\alpha$ line in the spectrum of HD~60\,848. Typical uncertainties are 0.25\,\AA\ for the EW and 0.01 for the V/R ratio. The insets in the top panel illustrate the first (18 December 2009, left inset) and last (14 March 2010, right inset) normalized spectra around the H$\alpha$ line as observed during the Mons campaign.\label{zoomHD60848}}
\end{figure}

Figure\,\ref{HD60848history} illustrates the long-term behaviour of the H$\alpha$, H$\beta$, and He\,{\sc i} $\lambda$\,5876 emissions in the spectrum of HD~60\,848. The first maximum of EW(H$\alpha$) and the subsequent decay have already been reported in Rauw et al.\ (\cite{ibvs}). Figure\,\ref{HD60848history} reveals another episode of maximum H$\alpha$ emission around HJD~2\,454\,500. By chance, the Mons campaign covers the decline of this episode and reveals a rather monotonic and fast drop of EW(H$\alpha$) (see Fig.\,\ref{zoomHD60848}). We have searched for periodicities in the EW(H$\alpha$) variations using both the Fourier-method for uneven sampling of Heck et al.\ (\cite{HMM}), modified by Gosset et al.\ (\cite{Gosset}), and the string-length method of Lafler \& Kinman (\cite{LK}), the latter being more appropriate for highly asymmetric variations. In this way, we find that the available observations suggest a recurrence timescale of about 2775\,days. 


\begin{figure}[h!]
\begin{center}
\resizebox{9cm}{!}{\includegraphics{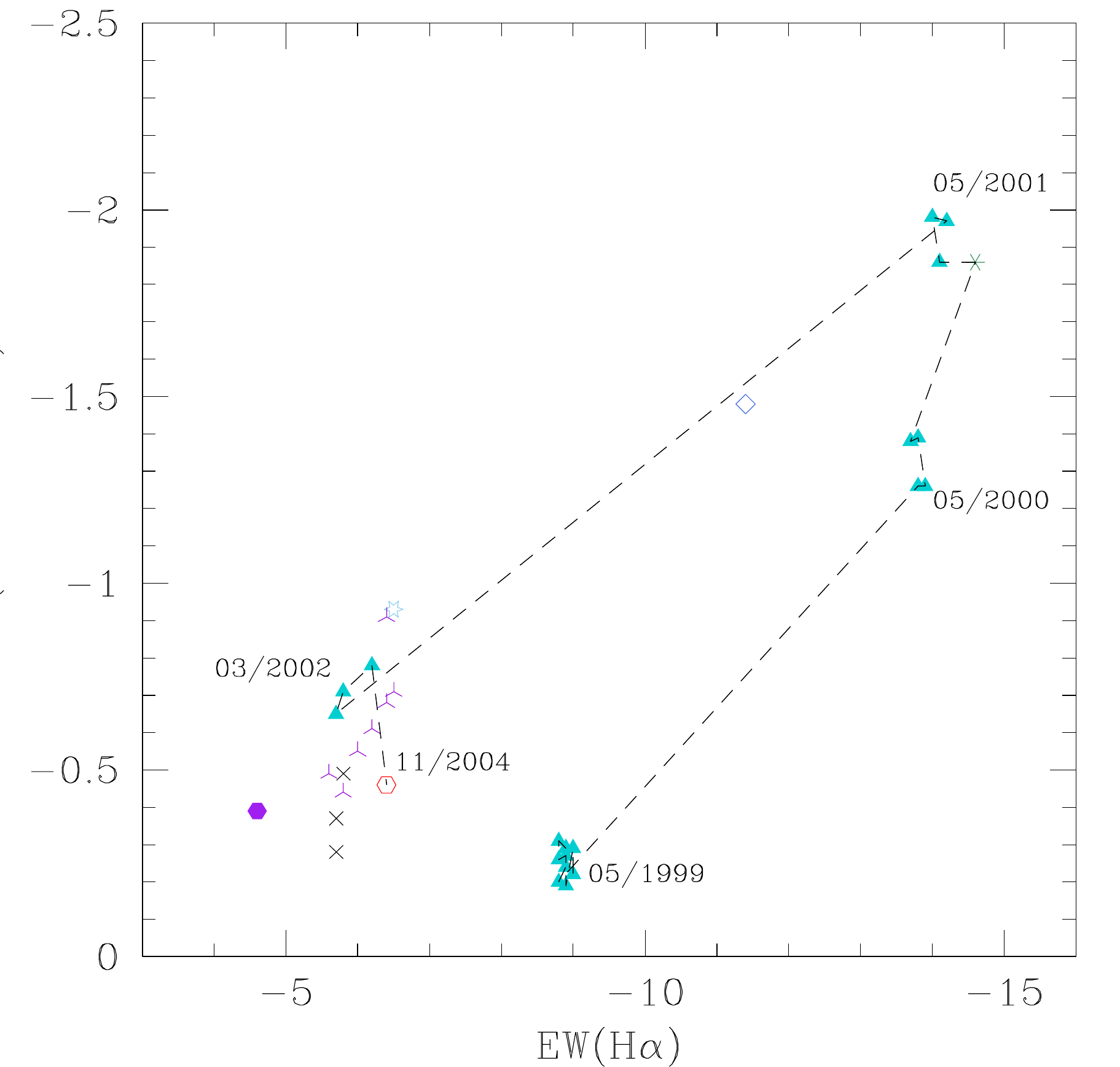}}
\end{center}
\caption{EW(He\,{\sc i} $\lambda$\,5876) as a function of EW(H$\alpha$) for HD~60\,848. The dashed line connects the data points taken during the high emission cycle between 1999 and 2004. The symbols have the same meaning as in Fig.\,\ref{HD60848history}.\label{HeI60848lag}}
\end{figure}

There is generally a good correspondence between the behaviour of the EWs of H$\alpha$ and He\,{\sc i} $\lambda$\,5876, although the variations in the latter lag a bit behind those of the former. This lag is clearly seen in Fig.\,\ref{HeI60848lag} where we show the EWs of both lines whenever they are available simultaneously.

\begin{figure}[h!]
\begin{center}
\resizebox{9cm}{!}{\includegraphics{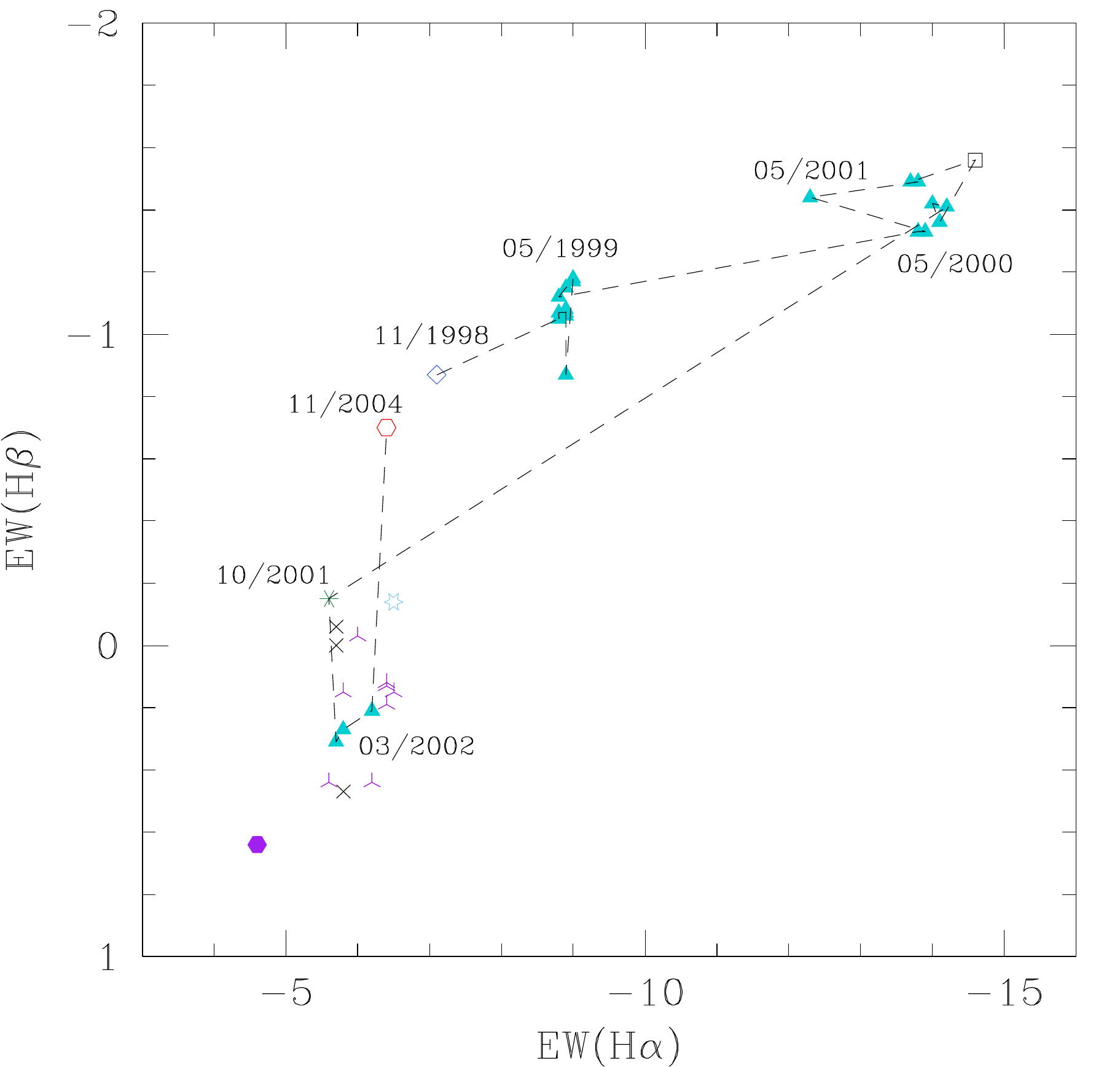}}
\end{center}
\caption{Same as Fig.\,\ref{HeI60848lag}, but for EW(H$\beta$) as a function of EW(H$\alpha$). \label{HD60848correl}}
\end{figure}

The situation is somewhat different for H$\beta$. For those observations where we have simultaneous access to the H$\alpha$ and H$\beta$ lines, we have again plotted the equivalent width of the former line versus that of the latter. The result is shown in Fig.\,\ref{HD60848correl}. When the emission lines have their lowest level, the H$\beta$ line is dominated by the photospheric absorption, leading to a positive equivalenth width, whilst H$\alpha$ is always seen in pure emission in our data. When the emission level increases, the equivalent widths of H$\beta$ increase at a faster rate than EW(H$\alpha$) (see Fig.\,\ref{HD60848correl}). This trend continues up to EW(H$\beta$) $\simeq -0.8$\,\AA, and EW(H$\alpha$) $\simeq -6.5$\,\AA. Beyond that point, the slope of the relation levels off and EW(H$\alpha$) now increases much faster than EW(H$\beta$). The decline of the emission level occurs very rapidly.   

The radial velocities of the He\,{\sc ii} absorption lines display some dispersion that might a priori be due to variability on relatively short timescales. For the He\,{\sc ii} $\lambda\lambda$\,4542, 4686, and 5412 lines, we obtain mean RVs of $35.4 \pm 11.4$, $24.0 \pm 5.7,$ and $25.7 \pm 7.5$\,km\,s$^{-1}$, respectively. Here the quoted dispersions correspond to standard deviations about the mean values. We have searched for periodicities using the methods introduced above, but no consistent periodicity was found. Boyajian et al.\ (\cite{Boyajian}) used a bisector method to measure the velocities of the emission lines of HD~60\,848 during an observing run in December 2000. They found possibly recurrent RV variations on timescales of three to four hours. These timescales are too short to be associated with orbital motion in a binary system, but could be due to non-radial pulsations, as suggested by McSwain et al.\ (\cite{McSwain}). Whilst our sampling is not suited to checking for the existence and stability of the periodicities reported by Boyajian et al.\ (\cite{Boyajian}), we note that the absorption lines indeed display line profile variability that might be at the origin of the RV variations. 

The V/R ratio of the H$\alpha$ emission remains very close to unity. The standard deviation about the mean value (1.00) of our 107 measurements amounts to 0.06. Our search for periodicities yields no significant result. The V/R ratios of H$\beta$ and He\,{\sc i} $\lambda$\,5876 behave similarly, although in these cases, the amplitude of variation increases when the overall level of emission goes down (which is as expected in view of the definition of V/R). The separation between the emission peaks of the H$\alpha$ line changes from about 210 -- 220\,km\,s$^{-1}$ at minimum emission to 150 -- 160\,km\,s$^{-1}$ at maximum emission level. In a Keplerian disk, where the separation between the peaks is interpreted as the value of $2\,V_{\rm Kep}\,\sin{i}$ at the outer rim of the disk (Horne \& Marsh \cite{HM}), such a variation indicates that the outer radius of the disk varied by a factor $\sim 1.9$ between minimum and maximum emission. We come back to this point in Sect.\,\ref{fits}.

\subsection{HD~45\,314}
\begin{figure}[h!]
\begin{center}
\resizebox{9cm}{!}{\includegraphics{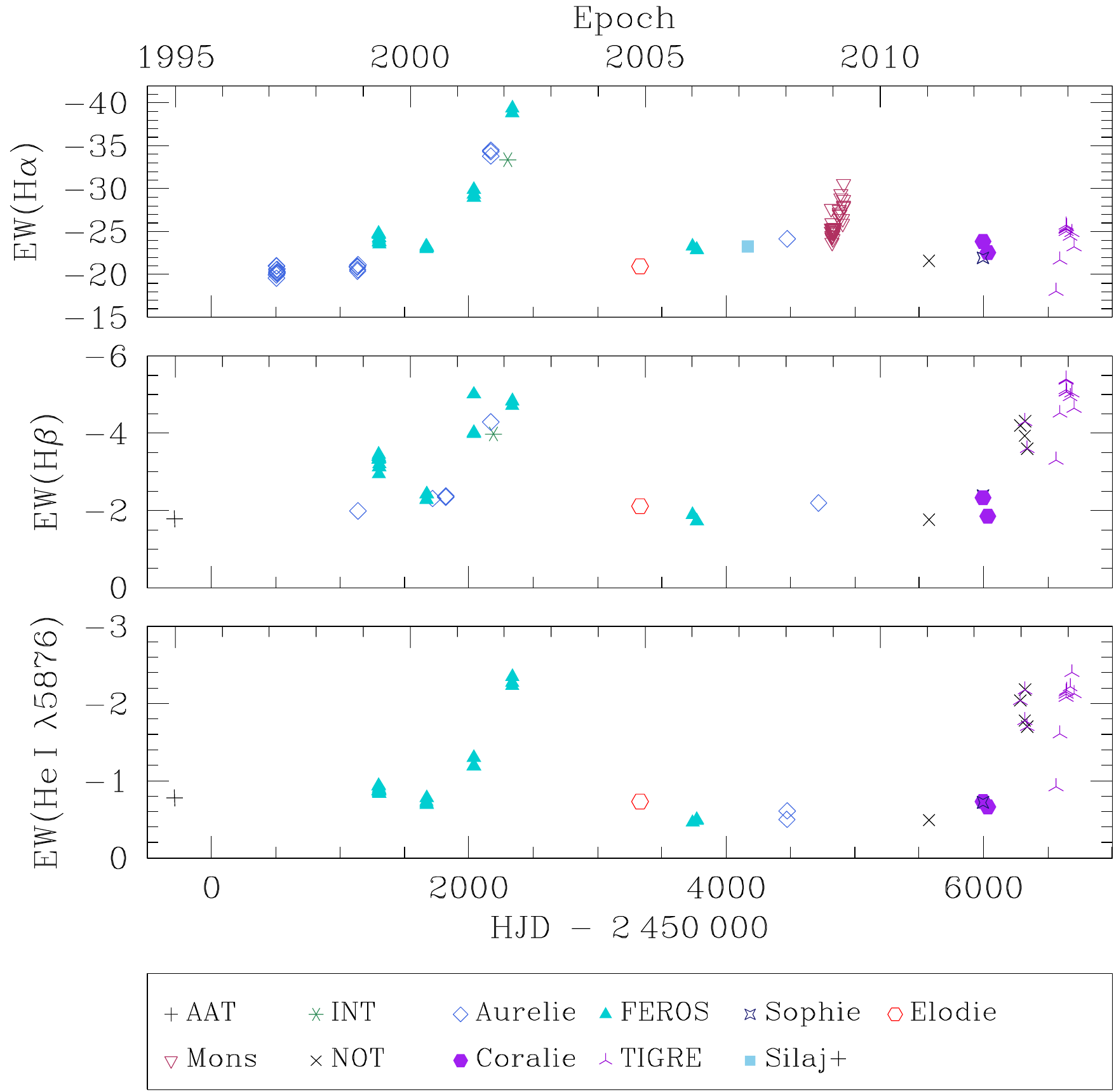}}
\end{center}
\caption{Same as Fig.\,\ref{HD60848history}, but for HD~45\,314. Typical uncertainties on the EWs are 0.50, 0.25, and 0.20\,\AA\ for the H$\alpha$, H$\beta$, and He\,{\sc i} $\lambda$\,5876 lines respectively. We have also included the EW(H$\alpha$) measurement of Silaj et al.\ (\cite{Silaj}). The second Coralie spectrum was taken simultaneously with the {\it XMM-Newton} observation discussed by Rauw et al.\ (\cite{Oeletter}). \label{HD45314history}}
\end{figure}

Figure\,\ref{HD45314history} shows the long-term behaviour of the equivalent widths of the H$\alpha$, H$\beta$, and He\,{\sc i} $\lambda$\,5876 emissions in the spectrum of HD~45\,314. These lines are always seen in pure emission in our data, even at those times of minimum global emission level. Copeland \& Heard (\cite{Copeland}) reported the absence of H$\beta$ and H$\gamma$ emission in 1962. Our data contain no such episode, at least not for H$\beta$.

\begin{figure}[h!]
\begin{center}
\resizebox{9cm}{!}{\includegraphics{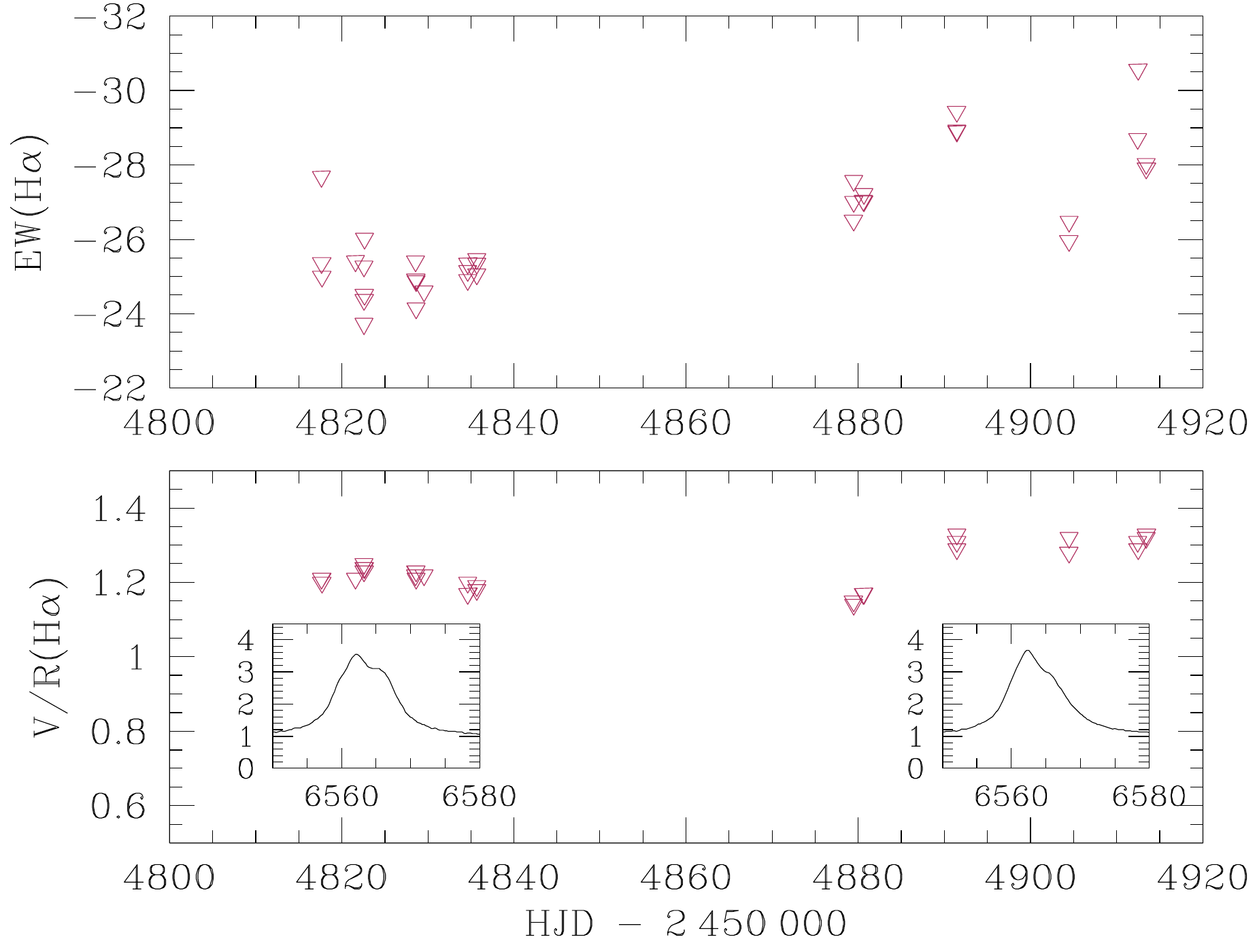}}
\end{center}
\caption{Same as Fig.\,\ref{zoomHD60848}, but for HD~45\,314. Typical uncertainties are 0.50\,\AA\ for EW and 0.01 for the V/R ratio. The insets in the lower panel illustrate the first (17 December 2009, left inset) and last (22 March 2010, right inset) normalized spectra around the H$\alpha$ line as observed during the Mons campaign.\label{zoomHD45314}}
\end{figure}

Figure\,\ref{HD45314history} reveals up to three episodes of high emission, although with different behaviours in the three lines. During the first maximum (HJD $\sim 2\,452\,500$), all three lines are in strong emission. The Mons campaign seems to cover another episode of slowly increasing H$\alpha$ emission (see Fig.\,\ref{zoomHD45314}), roughly 3200\,days after the first event. Unfortunately, we lack contemporaneous data in the other two lines. In our data a clear change in behaviour occurs for the third episode of strong emission after HJD~2\,456\,000. Whilst the H$\beta$ and He\,{\sc i} $\lambda$\,5876 lines both strongly increase, reaching their level of the first maximum, H$\alpha$ does not follow the trend and remains at a rather moderate level of emission\footnote{We have checked that none of the H$\alpha$ measurements used here is affected by saturation of the CCD. Some of the FIES @ NOT data suffer from saturation in this line, but they were discarded from our analysis.}. The behaviour of HD~45\,314 thus appears to change from one outburst to another.
\begin{figure}[h!]
\begin{center}
\resizebox{9cm}{!}{\includegraphics{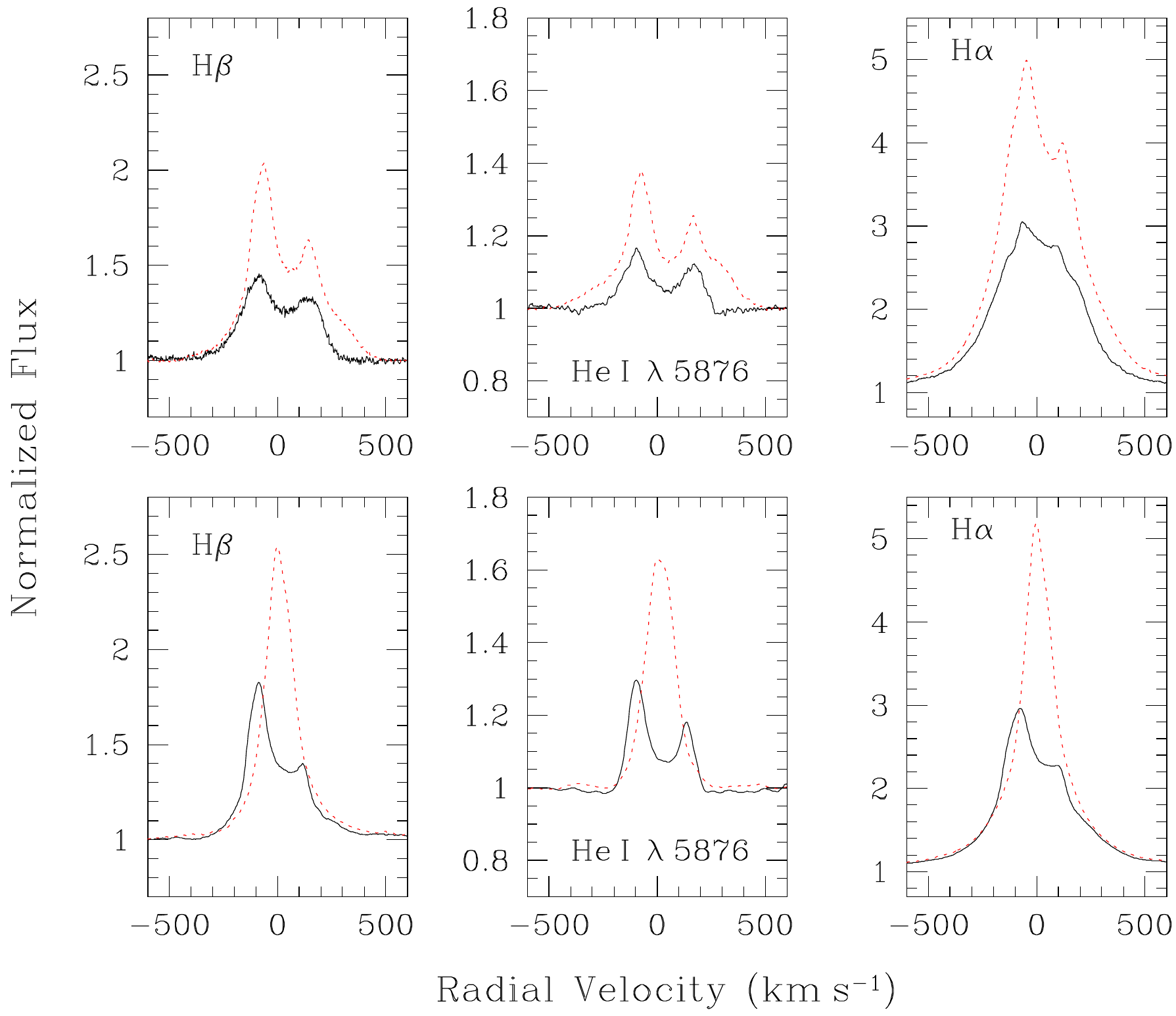}}
\end{center}
\caption{Line profile variations of HD~45\,314 between minimum and maximum emission at different epochs. The top row illustrates the variations in the spectrum between 7 May 2000 (black solid line, HJD~2\,451\,672.5) and 5 March 2002 (red dotted line, HJD~2\,452\,338.5). The bottom row yields the spectra as observed on 28 September 2013 (black solid line, HJD~2\,456\,564.0) and 20 December 2013 (red dotted line, HJD~2\,456\,646.9).\label{montage}}
\end{figure}

The general behaviour of HD~45\,314 during the 2013 event indeed seems different from the 2002 spectroscopic outburst. This is illustrated in Fig.\,\ref{montage}. In the 2002 event, the emission lines displayed similar morphologies at minimum and maximum, whilst these morphologies drastically changed between minimum and maximum in 2013. In December 2013, i.e.\ near maximum, all three lines were single-peaked. The H$\alpha$ emission remained single-peaked at least until February 2014. Whilst the maximum H$\alpha$ intensity reached roughly the same value as in the previous spectroscopic outburst, the big difference is the width of the line. In fact, in December 2013, the H$\alpha$ line at its maximum intensity is significantly narrower than previously.   

\begin{figure}[h!]
\begin{center}
\resizebox{9cm}{!}{\includegraphics{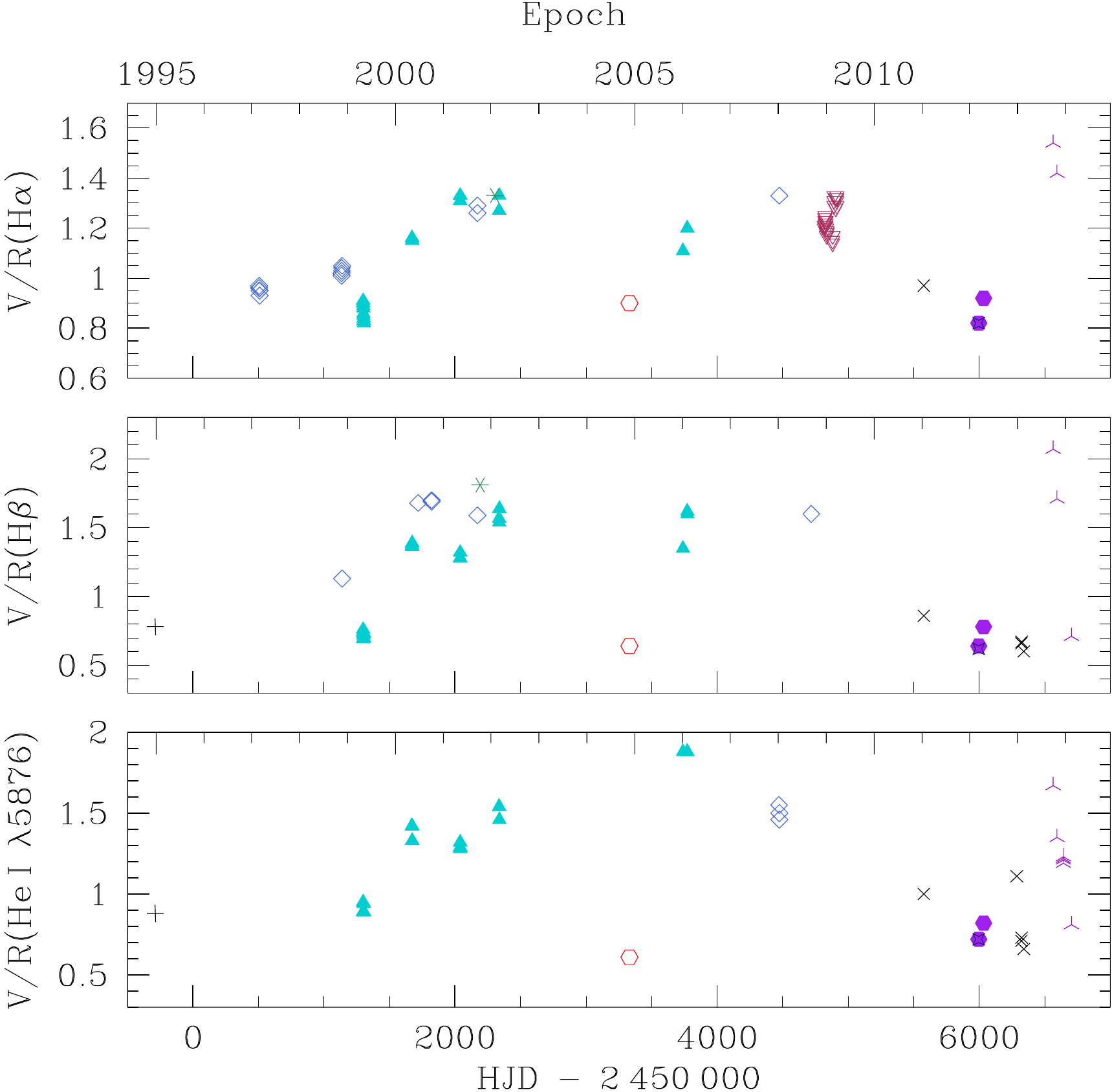}}
\end{center}
\caption{Variations in the V/R ratio of the H$\alpha$, H$\beta$, and He\,{\sc i} $\lambda$\,5876 emission lines in the spectrum of HD~45\,314. The symbols have the same meaning as in Fig.\,\ref{HD45314history}.\label{HD45314VR}}
\end{figure}

The V/R ratios of the emission lines in the spectrum of HD~45\,314 undergo large and rather slow variations between 0.82 and 1.56 for H$\alpha$, between 0.60 and 2.07 for H$\beta$, and between 0.66 and 1.88 for He\,{\sc i} $\lambda$\,5876 (see Fig.\,\ref{HD45314VR}). For those spectra where a single peak was observed, the measurements were discarded. Since the morphology of the H$\alpha$ line does not always allow the strengths of the peaks to be easily measured, we have also used the RV of the dominant peak as a proxy of the V/R variations of this line. A Fourier analysis of the time series of V/R measurements yields low-significance peaks at medium and long timescales. For the H$\beta$ line, the V/R variations are dominated by a timescale of $\sim 160$\,days. For the RVs of the H$\alpha$ peak and the V/R of He\,{\sc i} $\lambda$\,5876, we find a dominant timescale of $\sim 420$\,days in both cases. Finally, the V/R ratio of H$\alpha$ is modulated on a timescale of 2500\,days. The latter is also present in the periodograms of the V/R ratios of the H$\beta$ and He\,{\sc i} $\lambda$\,5876 lines, and of the RV of the dominant H$\alpha$ peak. We thus conclude that two possible timescales are 420 and 2500\,days, although their detections are not statistically significant.  
 
\begin{figure}[h!]
\begin{center}
\resizebox{9cm}{!}{\includegraphics{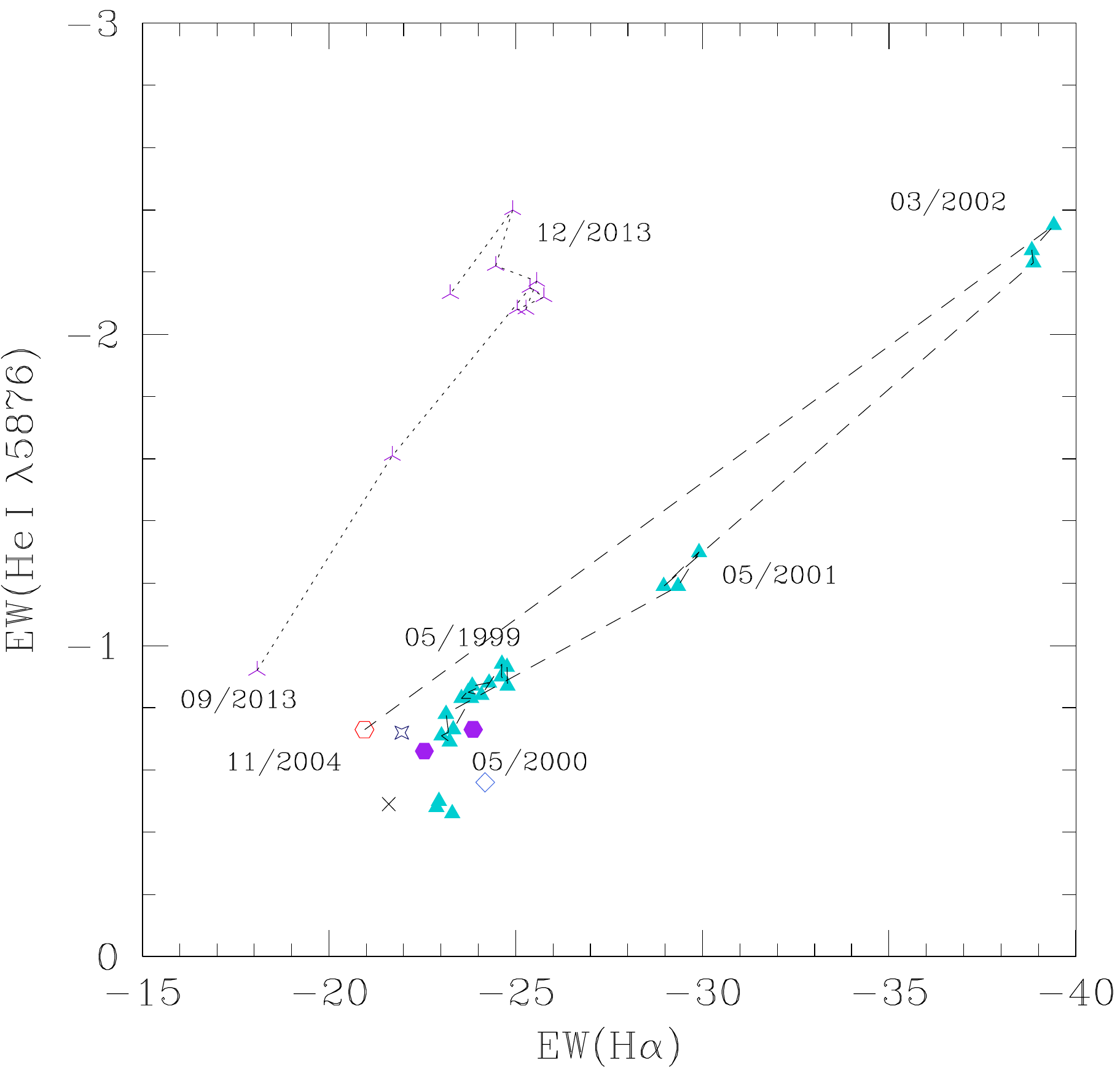}}
\end{center}
\caption{EW(He\,{\sc i} $\lambda$\,5876) as a function of EW(H$\alpha$) for HD~45\,314. The dashed line connects the data points taken around the 2002 spectroscopic outburst, whilst the dotted line connects the data from the 2013 outburst. The symbols have the same meaning as in Fig.\,\ref{HD45314history}.\label{HeI45314lag}}
\end{figure}

\begin{figure}[h!]
\begin{center}
\resizebox{9cm}{!}{\includegraphics{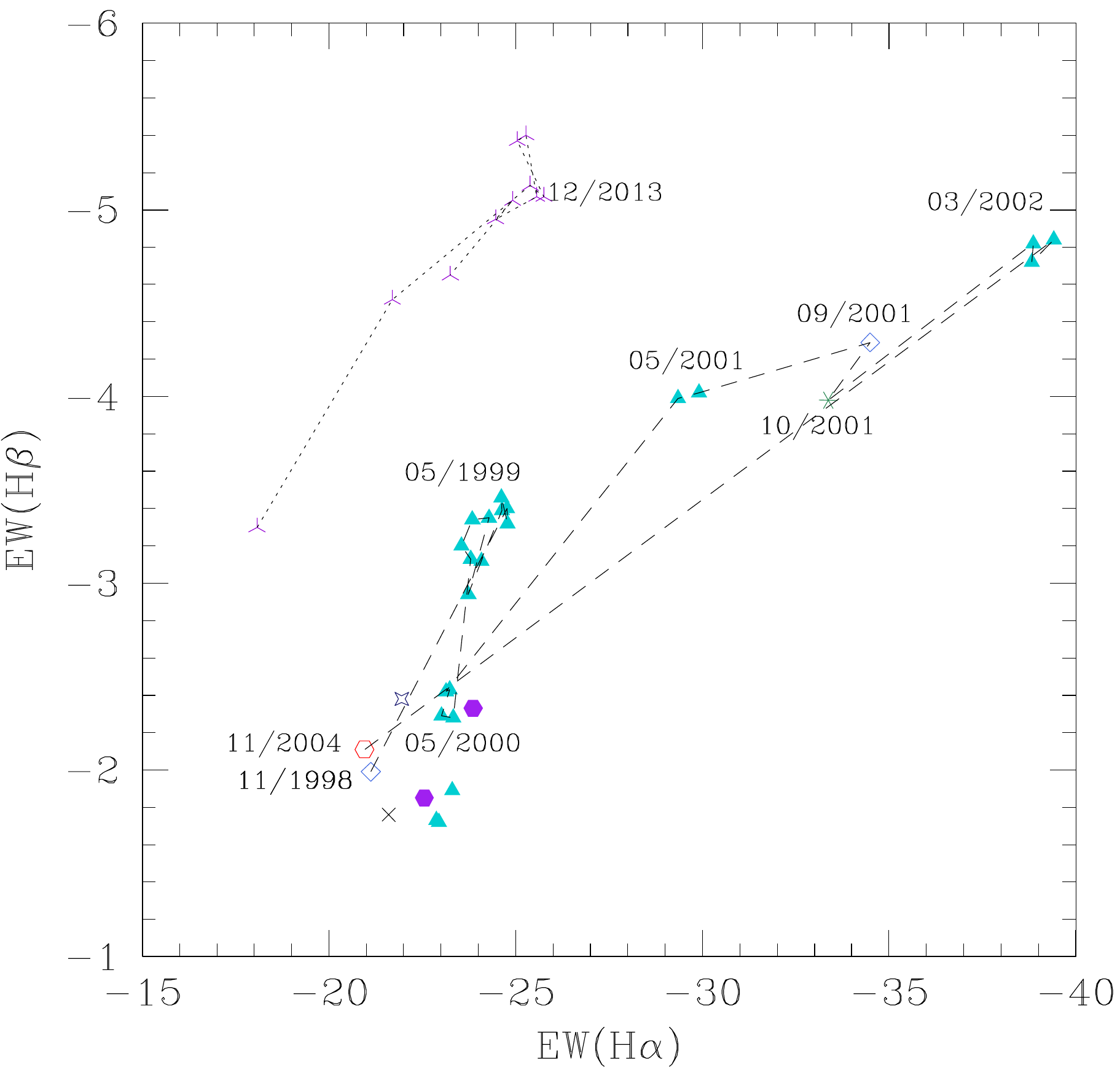}}
\end{center}
\caption{Same as Fig.\,\ref{HeI45314lag}, but for EW(H$\beta$) as a function of EW(H$\alpha$). The dashed line connects the data taken around the 2002 outburst, whilst the dotted line corresponds to the 2013 - 2014 event.\label{HD45314correl}}
\end{figure}

Unlike what we found for HD~60\,848, in the case of HD~45\,314, we observe no evidence for a systematic time delay between the variations in the EWs of H$\alpha$ and He\,{\sc i} $\lambda$\,5876 (see Fig.\,\ref{HeI45314lag}): the EWs of both lines seem well correlated during the outbursts though with a shift between the 2002 and 2013 events. In contrast, the relationship between EW(H$\alpha$) and EW(H$\beta$) is more complex (see Fig.\,\ref{HD45314correl}). At first sight the behaviour during the 2002 spectroscopic outburst is similar to what was seen in HD~60\,848 (compare with Fig.\,\ref{HD60848correl}). However, as for the plot of EW(He\,{\sc i} $\lambda$\,5876) versus EW(H$\alpha$), the 2013 event is clearly offset with respect to this behaviour. This is due to the significantly lower level of H$\alpha$ emission in 2013.

The RVs of the He\,{\sc ii} $\lambda$\,4686 absorption line yield a mean value of $-1.8 \pm 14.3$\,km\,s$^{-1}$. As for HD~60\,848, we have searched for periodicities using the Fourier and string-length methods introduced above. No significant periodicity is found in our RV time series. Boyajian et al.\ (\cite{Boyajian}) found a significant RV shift between their data of HD~45\,314 taken in October and December 2000. As for HD~60\,848, the RVs of Boyajian et al.\ (\cite{Boyajian}) refer to bisector velocities of the emission lines, and the shift reported by these authors could thus actually be due to changes in the shape of the emission line profiles.

\section{Discussion}
\subsection{Fitting the H$\alpha$ profiles \label{fits}}
Assuming that both Oe stars feature a Keplerian decretion disk, we can attempt to quantify the properties of these disks at different epochs. For this purpose, we designed a simple model that allows the emission profile from an optically-thick disk to be computed. Our numerical code is based on the approach of Hummel \& Vrancken (\cite{HV}), which in turn follows the formalism of Horne \& Marsh (\cite{HM}) for computing accretion disks in cataclysmic variables. A very similar approach was adopted by Grundstrom \& Gies (\cite{GG}) and Catanzaro (\cite{Catanzaro}) to compute synthetic Be emission line profiles and compare them with actual observations. 

The disk is assumed to be in pure Keplerian rotation and to extend from the stellar surface out to an outer radius $R_{\rm disk}$. The particle number density in the disk is described as a power law of index $\alpha$ in radius ($r$) and a Gaussian profile in elevation ($z$) above the disk plane:
\begin{equation} 
n(r,z) = n_0\,\left(\frac{r}{R_*}\right)^{-\alpha}\,\exp{\left[-\frac{1}{2}\left(\frac{z}{H(r)}\right)^2\right]}
\end{equation}
where $H(r)$ is the local scale height (see equations 2 and 4 of Hummel \& Vrancken \cite{HV}). At each position in the disk, the line-of-sight optical depth ($\tau$) is computed accounting for the line profile width due to the thermal velocity and the shear broadening (see equations 7, 8, and 9 of Hummel \& Vrancken \cite{HV}). 
The specific intensity along the line of sight is computed according to
\begin{equation}
I_{\lambda} = B_{\lambda}(T_{\rm disk})\,[1 - \exp{(-\tau)}] + I_{\lambda}^* \,\exp{(-\tau)}
,\end{equation}
where we assume an isothermal disk and adopt the Planck function at $T_{\rm disk} = 0.6\,T_{\rm eff}$ as the source function. Occultation of the stellar surface by the disk is accounted for by the $I_{\lambda}^*\,\exp{(-\tau)}$ term, and occultation of the disk by the stellar body is also taken into account. 
Since we assume an axisymmetric disk, the synthetic line profiles are by definition symmetric about the central wavelength. We note that electron scattering is not accounted for in our model. 

For our computations, we have adopted $T_{\rm eff} = 32500$\,K, $R_* = 8$\,R$_{\odot}$, and $M_* = 20$\,M$_{\odot}$ which are the mean values of the stellar parameters of HD~45\,314 and HD~60\,848 as quoted by Vink et al.\ (\cite{Vink}). With these numbers fixed, there remain four free parameters in the model: the orbital inclination $i$, the outer disk radius $R_{\rm disk}$, the power-law index $\alpha$, and the particle number density at the inner edge of the disk $n_0$. 

To account for the underlying photospheric absorption line, we adopted the effective temperatures ($31500$ and $33400$\,K for HD~45\,314 and HD~60\,848, respectively) and surface gravity ($\log{g} = 3.92$ for both stars) as quoted by Vink et al.\ (\cite{Vink}). We then interpolated in the Ostar2002 grid of non-LTE plane-parallel TLUSTY models (Lanz \& Hubeny \cite{LH}) for solar metallicity. The resulting spectrum was broadened with the projected rotational velocities inferred above to obtain the photospheric spectra. 

In our comparison with the actual observations, we focused on the H$\alpha$ line. First we computed a grid of synthetic line profiles over a wide range of parameters ($i$ from 15 to $75^{\circ}$ with a step of $5^{\circ}$, $R_{\rm disk}$ between 2 and 30\,$R_*$ by stpdf of 1\,$R_*$, $\alpha$ between 1.5 and 3.5 with a step of 0.5, and $n_0$ varying between 16 and 626\,cm$^{-3}$ with a step of 16\,cm$^{-3}$). For each star, we considered the observed H$\alpha$ profile at several epochs: 1999, 2000, and 2002. For each of these epochs, we averaged all the FEROS @ 1.5m\,ESO data that are available. We also included the average HEROS @ TIGRE spectrum of HD~60\,848 as observed in 2013 -- 2014. For HD~45\,314, we further considered the mean spectrum of the 2011 FIES @ NOT dataset and the HEROS @ TIGRE spectrum taken on 20 December 2013 (see Fig.\,\ref{montage}).

\begin{figure}[h!]
\begin{center}
\resizebox{9cm}{!}{\includegraphics{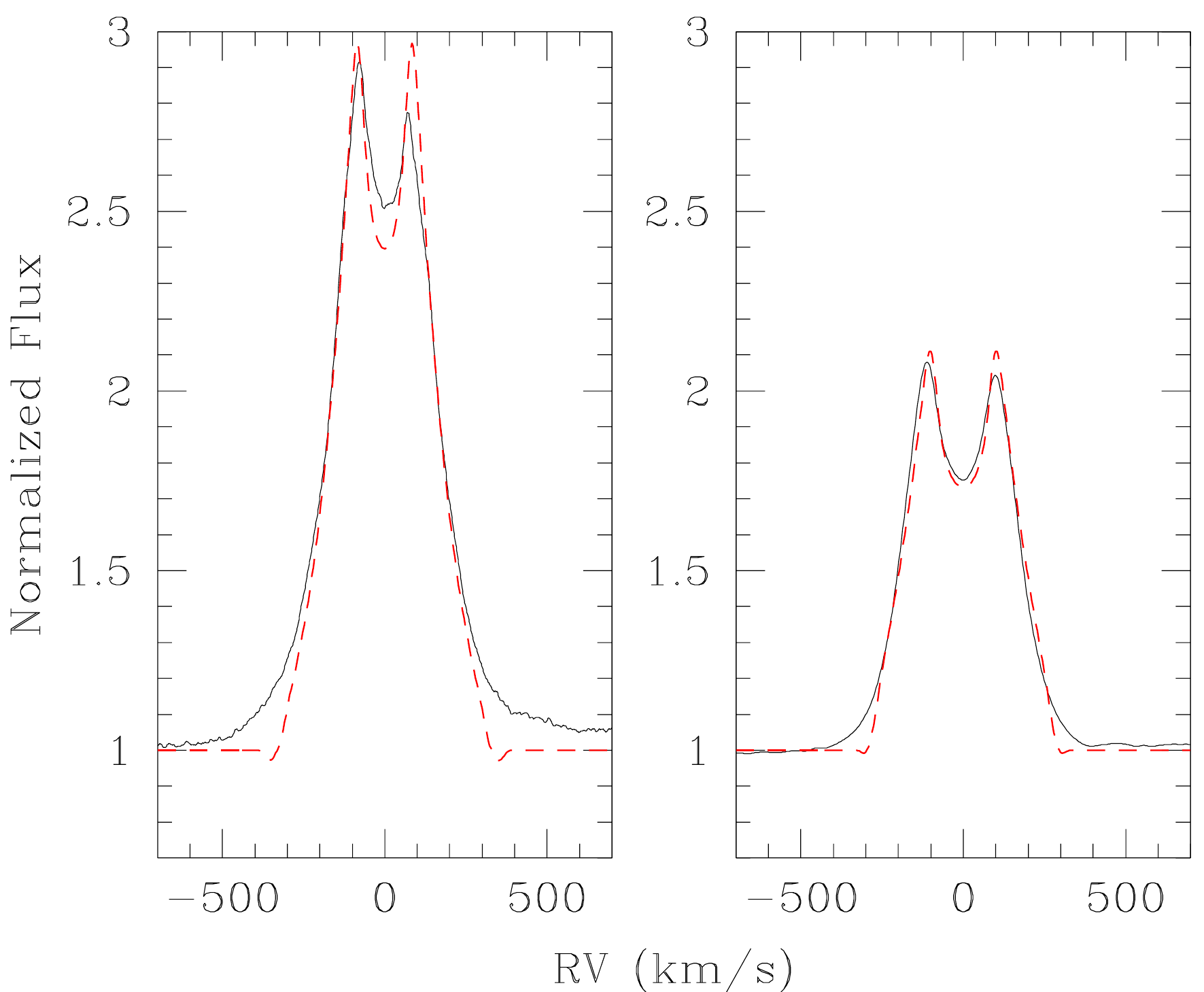}}
\end{center}
\caption{Fits of the H$\alpha$ line of HD~60\,848 as observed in May 2000 (left) and the winter season 2013-2014 (right) with our disk model. The model is shown by the dashed line. The corresponding model parameters are given in Table\,\ref{Tablefit}. The failure to reproduce the line wings is probably because our synthetic models neglect electron scattering. \label{fit60848}}
\end{figure}

HD~60\,848 presents only small-amplitude V/R variations and thus displays quite symmetric line profiles. These properties make this star well suited for application to our model. Fitting our model to the H$\alpha$ line profiles from all four epochs indeed yields good results. As an example, Fig.\,\ref{fit60848} displays the fits for two epochs that roughly correspond to the most extreme values of EW(H$\alpha$). The best-fit model parameters are listed in Table\,\ref{Tablefit}. As can be seen, all epochs yield an inclination of the disk near $30^{\circ}$. The inclination inferred from the 1999 data is slightly larger than at the other epochs, but this may not be significant. The best-fit inclination yields an equatorial rotational velocity of 460\,km\,s$^{-1}$, i.e.\ about 67\% of the critical velocity for the stellar parameters adopted here. However, as pointed out above, this estimate could be biased by gravity darkening effects, because we have derived the $v\,\sin{i}$ value from the He\,{\sc ii} $\lambda$\,4686 line, which is likely to arise from the polar regions where the rotation is slower.

All observations are consistent with a density-law exponent $\alpha \simeq 3.0$. The outer radius of the disk varies from about 12\,$R_*$ at minimum emission state to 28\,$R_*$ at the intermediate emission state observed in May 1999. Surprisingly, the spectrum with the strongest H$\alpha$ emission (May 2000) does not correspond to the largest disk radius, but rather corresponds to the highest density $n_0$. We repeated the fitting by fixing the inclination to $30^{\circ}$ and $\alpha$ to 3.0. The main changes with respect to the previous results are a reduction of the disk radius for the May 1999 epoch to $(21.5 \pm 1.7)$\,$R_*$ associated with a 17\% reduction of $n_0$. The disk radius in May 1999 is now comparable to the value inferred for the May 2000 epoch (20.5\,$R_*$). Our results thus suggest that the high state in HD~60\,848 corresponds to an increase in disk radius followed by an increase in the density in the disk.   

The best-fit parameters in Table\,\ref{Tablefit} are broadly consistent with the properties inferred for classical Be stars. Indeed, $n_0$ values inferred by Hummel \& Vrancken (\cite{HV}) span a wide range from 33 to 1005\,cm$^{-3}$, and our results are well inside this range. The density-law exponents of Be disks are close to $3$ (with a range from 1.6 to 3.6, Hummel \& Vrancken \cite{HV}) as we also find for HD~60\,848. The only parameter that stands out in the case of HD~60\,848 is the outer radius of the disk. Be disks are usually found to be smaller with most objects having $R_{\rm disk}$ less than 12\,$R_*$ (Grundstrom \& Gies \cite{GG}, Catanzaro \cite{Catanzaro}) and only a few objects seem to have disks as large as 18.5\,$R_*$ (Hummel \& Vrancken \cite{HV}).  
\begin{table}[t!]
\begin{center}
  \caption{Best-fit parameters of the H$\alpha$ line profile of HD~60\,848. The quoted uncertainties correspond to the 90\% confidence interval.\label{Tablefit}}
\begin{tabular}{c c c c c}
\hline
Epoch & $i$ ($^{\circ}$) & $R_{\rm disk}$ ($R_*$) & $n_0$ (cm$^{-3}$) & $\alpha$ \\
\hline
May 1999 & $36.7 \pm 3.3$ & $27.9 \pm 1.6$ & $420 \pm 117$ & $2.9 \pm 0.2$ \\
May 2000 & $30.0 \pm 3.0$ & $20.4 \pm 1.1$ & $585 \pm 31$ & $3.0 \pm 0.2$ \\
Mar.\ 2002 & $28.2 \pm 3.4$ & $14.4 \pm 2.7$ & $270 \pm 94$ & $3.1 \pm 0.3$ \\
2013-14 & $29.2 \pm 3.4$ & $12.6 \pm 2.3$ & $199 \pm 92$ & $2.8 \pm 0.4$ \\
\hline
\end{tabular}
\end{center}
\end{table}
 
\begin{figure}[h!]
\begin{center}
\resizebox{9cm}{!}{\includegraphics{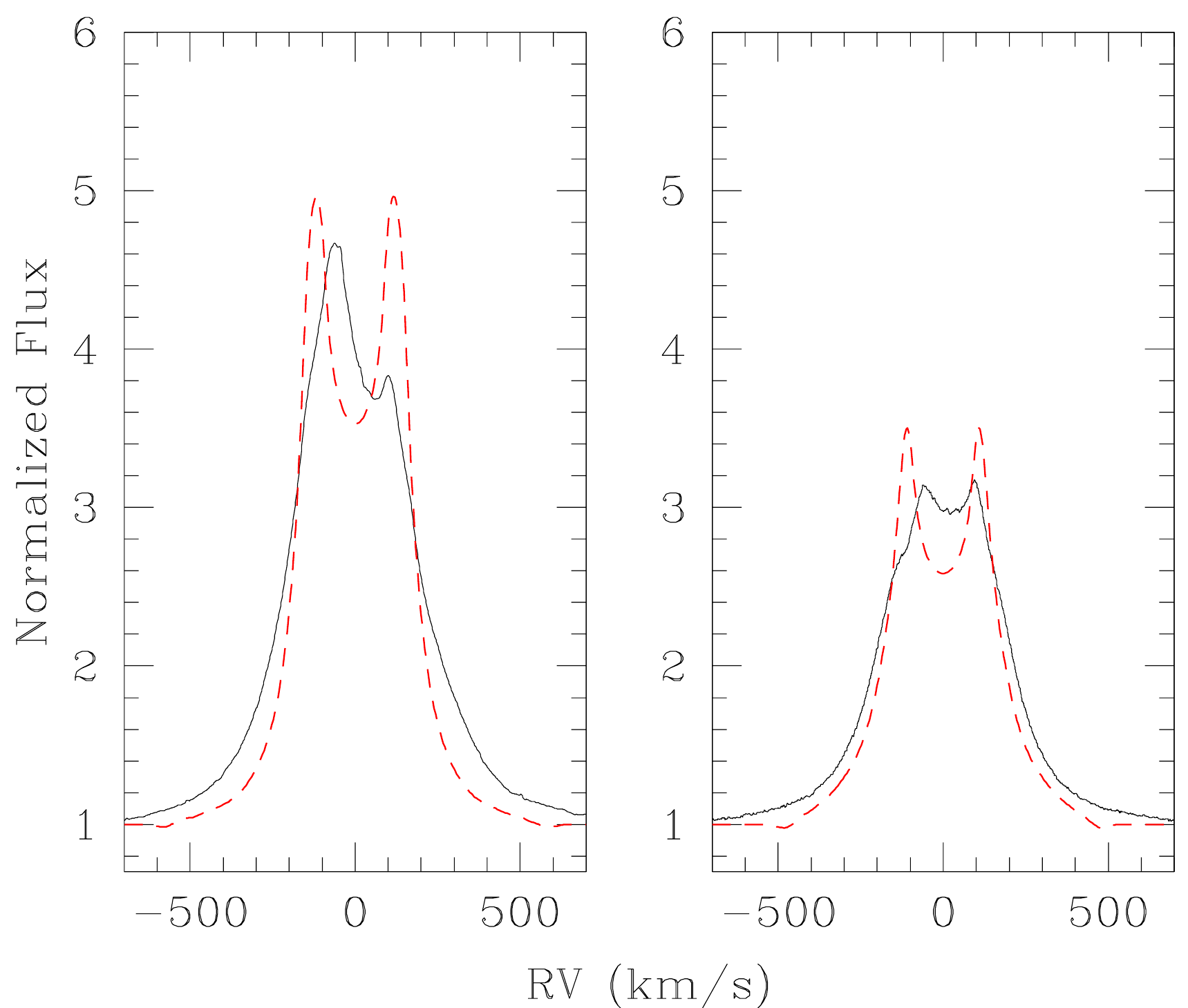}}
\end{center}
\caption{Fits of the H$\alpha$ line of HD~45\,314 as observed in March 2002 (left) and January 2011 (right) with our disk model. The model is shown by the dashed line.\label{fit45314}}
\end{figure}

The situation is more complex for HD~45\,314. Indeed, the observed line profiles of this star are more asymmetric and display significant V/R variations. Moreover, the line profile variations not only concern the relative strengths of the peaks, but also result in bumps in the emission wings. Furthermore, the observed profiles display broad emission wings that suggest a velocity at the inner rim of the disk that exceeds the velocity at the stellar radius $\sqrt{\frac{G\,M_*}{R_*}}$ calculated from the adopted values of $M_* = 20$\,M$_{\odot}$ and $R_* = 8$\,R$_{\odot}$. All these facts suggest that the velocity field in the emission volume around HD~45\,314 is more complex than the simple smooth Keplerian disk that we assume in our model. Unsurprisingly, our model fails to correctly reproduce the observations. The least disappointing results are obtained for the FIES@NOT spectra taken in 2011 (see Fig.\,\ref{fit45314}). Still, this fit has much lower quality than what we achieved for HD~60\,848. For the May 1999, May 2000, and January 2011 observations, we obtain a `best-fit'inclination of $45^{\circ}$ and a disk radius of $\simeq 22.5$\,$R_*$. The index of the density law is found to be $2.5$. However, this apparent consistency between the parameters does not hold for observations taken when the star is in high state. Indeed, the March 2002 data yield a `best-fit' inclination of $60^{\circ}$, whilst the 2013 spectra suggest $25^{\circ}$. This situation essentially reflects our failure to achieve a decent fit of the observed profiles with our model.   

The formal quality of the fit to the broad wings\footnote{These broad wings are not artefacts due to normalization errors of the echelle spectra. In fact, the same features are seen in the Aur\'elie data, which are much easier to normalize.} of the 2002 spectra could be improved by artificially increasing the $M_*/R_*$ ratio by a factor $\sim 4$. Such a high ratio is, however, at odds with the typical parameters of Population I O-type stars, and the agreement would not be any better if assuming typical masses and radii of subdwarf B-stars (Heber \cite{Heber}). 

As pointed out above, our failure to correctly reproduce the line profiles probably points towards a significantly more complex velocity field than assumed in our model. Whilst the observed asymmetries of the line profiles can be explained, at least to first order, assuming a sinusoidal modulation of the disk density, the changes in line widths and in the inferred disk inclination are more challenging. Genuine inclination changes could stem from the torque from a neutron star companion if the disk were misaligned with the neutron star's orbit. Martin et al.\ (\cite{Martin}) show that in such a configuration, the disk could be truncated inside the neutron star's orbit, preventing Type II X-ray outbursts. Such a disk would be warped and would precess on a timescale that is much longer than the neutron star's orbit. However, we currently lack clear evidence of such a neutron star companion in HD~45\,314 because there is no convincing signature of RV variations due to binarity, and the X-ray properties of this star differ from those of classical high-mass X-ray binaries. Moreover, given the poor quality of some of the line profile fits, the evidence of inclination changes is rather weak. 

Finally, we have tested whether a non-Keplerian velocity law could improve the situation. For this purpose, we computed models with a velocity law $v_{\rm rot}(r) = \sqrt{\frac{G\,M_*}{R_*}}\,\left(\frac{r}{R_*}\right)^{-\beta}$. In this relation $\beta = 0.5$ corresponds to a Keplerian velocity field, whilst $\beta = 1$ corresponds to conservation of angular momentum. Neither $\beta = 0$ nor $\beta = 1$ allow improvement in the quality of the fit of the observed profiles.    
 
\subsection{The time dependence of the spectra}
In the case of the B2\,IV-Ve star $\omega$\,CMa, \v{S}tefl et al.\ (\cite{Stefl}) report a time delay of several months between the increase in the continuum flux and higher order Balmer lines, on the one hand, and the increase in H$\alpha$, on the other hand. These authors note further that the path from quiescence to high-emission state is distinct from the return path in the Balmer decrement diagrams (H$\gamma$/H$\beta$ versus H$\alpha$/H$\beta$). As a result, the star describes a hysteresis-like path that \v{S}tefl et al.\ (\cite{Stefl}) interpret in terms of the build-up and decay of the disk, both of which progress radially outwards. In this scenario, the mass-weighted mean radius of the disk would be smaller during build-up than during decay. 

The time dependence of the equivalent widths of the H$\alpha$ and H$\beta$ emission lines in Fig.\,\ref{HD60848correl} indicates that in the case of HD~60\,848, H$\alpha$ also seems to lag behind H$\beta$, at least during the early stages of the outburst. At first sight this is reminiscent of the situation of $\omega$\,CMa. That the increase of the H$\beta$ line levels off whilst H$\alpha$ continues to increase could indicate that the excitation of the $n=3$ level of hydrogen in the disk continues to progress outwards, once the region of the disk where the $n=4$ level is excited has reached its maximum extension. Alternatively, when the disk becomes very dense, parts of it could become cooler leading to a reduction in the relative population of the upper levels of hydrogen and thus a slower increase in H$\beta$. However, these scenarios are contradicted by Fig.\,\ref{HeI60848lag}, which indicates that He\,{\sc i} $\lambda$\,5876 lags behind the variations in H$\alpha$: the EW of He\,{\sc i} $\lambda$\,5876 continues increasing once EW(H$\alpha$) has reached its maximum, even though the excitation of the upper level of He\,{\sc i} $\lambda$\,5876 requires almost twice as much energy as the excitation of the $n=3$ level of hydrogen. 

The situation is again different for HD~45\,314. No hysteresis-like loops were found, but the star changes its behaviour from one epoch to the next. Whether this is related to the status of HD~45\,314 as a $\gamma$~Cas analogue (Rauw et al.\ \cite{Oeletter}) is not clear. At least the transition between double- and single-peaked emission lines is not restricted to $\gamma$~Cas analogues because such transitions have been reported for other Be stars (e.g.\ Hanuschick \cite{Hanuschick}). To our knowledge, apart from $\gamma$~Cas (B0.5\,IVpe) itself (Miroshnichenko et al.\ \cite{Mirosh}, Pollmann et al.\ \cite{Pollmann}), none of the currently known $\gamma$~Cas analogues have been monitored spectroscopically on long timescales. In their long-term study of $\gamma$~Cas, Miroshnichenko et al.\ (\cite{Mirosh}) reported V/R variations with cycles of progressively increasing length ($\sim$ 4, 5, 7, and 10\,years). The V/R ratios of the H$\alpha$ and non-hydrogen lines in $\gamma$~Cas were found to vary non-simultaneously, with the H$\alpha$ cycle being longer. In HD~45\,314, we do not find evidence of any such effect, since the H$\alpha$, H$\beta,$ and He\,{\sc i} $\lambda$\,5876 lines all display similar timescales in their V/R variations. 
  
\section{Summary}
In this paper we have reported analysis of an extensive set of optical spectra of the two Oe stars, HD~45\,314 and HD~60\,848. The morphology of the H$\alpha$ emission of HD~60\,848 is explained well by a Keplerian disk around the star, despite previous reports of a lack of a spectropolarimetric signature for such a disk (Vink et al.\ \cite{Vink}). The emission lines of HD~60\,848 undergo strong variations in their strength, with H$\beta$ increasing first, followed by H$\alpha,$ and finally by He\,{\sc i} $\lambda$\,5876. This situation cannot be explained by simple considerations of the progressive excitation of the disk, and the origin of the time lags is currently unclear. 

Whilst HD~45\,314 is the only Oe star in the sample of Vink et al.\ (\cite{Vink}) where a clear depolarization effect across the H$\alpha$ line was found in the linear spectropolarimetry, our simple Keplerian disk model fails to reproduce the observed line profiles of this star. This star furthermore exhibits strong variations in the morphology and strength of its emission lines. Moreover, we found a change in behaviour between previous high-emission states and the most recent event (2013 -- 2014). Currently, the reasons for this strange behaviour and the connections with the fact that this star is a $\gamma$~Cas analogue are unclear. Our X-ray observation of HD~45\,314 was obtained when the star was in a low state. If the $\gamma$~Cas status is directly related to phenomena in the disk, as suggested by Smith et al.\ (\cite{Smith1}), one would expect the X-ray emission to be different during a high state. We will thus continue our optical monitoring of this star with the goal of triggering an X-ray observation during its next outburst.        

In the sample of Vink et al.\ (\cite{Vink}), four Oe stars display a strong H$\alpha$ emission (HD~45\,314, HD~60\,848, HD~120\,678, and HD~155\,806), but only HD~45\,314 displays a strong depolarization effect. One can thus speculate whether the positive detection of HD~45\,314 might stem from one of its peculiarities: the strong V/R variations, hinting at a precessing density wave, or its status as a $\gamma$~Cas analogue. The first possibility is unlikely because the spectra of HD~120\,678 (O8\,III:nep) collected by Gamen et al.\ (\cite{Gamen}) over five years also reveal strong V/R variations\footnote{These authors also report a shell-like event in HD~120\,678 similar to shell episodes seen in some Be stars.} for this star, although it shows no strong depolarization effect. Concerning the second possibility, we note that whilst nothing is known about the X-ray emission of HD~120\,678, the X-ray emission of HD~155\,806 and HD~60\,848 are typical of normal OB stars (Naz\'e et al.\ \cite{Naze}, Rauw et al.\ \cite{Oeletter}) and HD~45\,314 is thus currently the only known $\gamma$~Cas analogue among the Vink et al.\ (\cite{Vink}) sample. Therefore, we cannot rule out the possibility that the strong depolarization effect in HD~45\,314 might be related to its status as a $\gamma$~Cas analogue rather than to the mere presence of a disk.

\section*{Acknowledgements}
The professional authors of this paper are grateful to the amateur astronomers of the Mons team, who invested personal time, money, and enthusiasm in this project. We thank the Director of the IAC, Prof.\ F.\ Sanchez, and Dr.\ M.\ Serra for allocation of observing time and assistance at the Mons telescope. This work makes use of data obtained from the Isaac Newton Group of telescopes and Anglo Australian Telescope data archives maintained at the Institute of Astronomy Cambridge, as well as data retrieved from the ELODIE archive at the Observatoire de Haute Provence and the archives of the Nordic Optical Telescope. We also used data collected at the European Southern Observatory (La Silla \& Cerro Paranal, Chile), at the Observatoire de Haute Provence, and with the TIGRE at La Luz Observatory (Guanajuato, Mexico). The TIGRE is funded and operated by the universities of Hamburg, Guanajuato, and Li\`ege. The Li\`ege team acknowledges support from the Fonds de Recherche Scientifique (FRS/FNRS), through the XMM/INTEGRAL and GAIA-DPAC PRODEX contract, as well as by the Communaut\'e Fran\c caise de Belgique - Action de recherche concert\'ee - Acad\'emie Wallonie - Europe. AFJM is grateful for financial assistance from NSERC (Canada) and FRQNT (Quebec). JHK acknowledges financial support to the DAGAL network from the People Programme (Marie Curie Actions) of the EU FP7/2007-2013/ under REA grant agreement number PITN-GA-2011-289313, and from the Spanish MINECO under grant number AYA2013-41243-P. The authors are grateful to Douglas Gies, Thomas Rivinius, and Myron Smith for discussion of some of the results of this work.

%% References
%%
%% Following citation commands can be used in the body text:
%% Usage of \cite is as follows:
%%   \cite{key}         ==>>  [#]
%%   \cite[chap. 2]{key} ==>> [#, chap. 2]
%%

%% References with bibTeX database:

\bibliographystyle{elsarticle-harv}
%\bibliography{<your-bib-database>}

%% Authors are advised to submit their bibtex database files. They are
%% requested to list a bibtex style file in the manuscript if they do
%% not want to use elsarticle-num.bst.

%% References without bibTeX database:

% \begin{thebibliography}{00}

%% \bibitem must have the following form:
%%   \bibitem{key}...
%%

% \bibitem{}

% \end{thebibliography}

\end{document}